\documentclass[floatfix,aps,pra,superscriptaddress,twocolumn,longbibliography]{revtex4-2}

\usepackage[english]{babel}

\usepackage{graphicx}
\usepackage[colorlinks=true, allcolors=blue]{hyperref}
\usepackage{physics}
\usepackage{amsmath,amsfonts,amssymb,amsthm,comment}
\usepackage{enumitem}
\usepackage[utf8]{inputenc}
\usepackage{lipsum}
\usepackage{array,hhline,makecell}
\usepackage{ulem}
\usepackage{tabularx}
\newcolumntype{Y}{>{\centering\arraybackslash}X}
\usepackage[dvipsnames]{xcolor}

\usepackage{natbib}
\usepackage{url}
\hypersetup{colorlinks=true, urlcolor=blue, linkcolor=blue, citecolor=blue}

\newcommand{\be}{\begin{eqnarray}}
\newcommand{\ee}{\end{eqnarray}}

\begin{document}

\title{A  thermodynamic analysis of nonsecular master equations, beyond strict energy conservation and Gibbs equilibrium state}

\author{Mohamed Boubakour}
\email{mohamed.boubakour@univ-lorraine.fr}
\author{Talia Szikman}
\author{Cyril Elouard}
\affiliation{Université de Lorraine, CNRS, LPCT, F-54000 Nancy, France}

\date{\today}

\begin{abstract}
We present a systematic thermodynamic analysis of nonsecular GKLS master equations. When some energy transitions in the system are comparable with the dissipative rate, the secular approximation is not valid, and the positivity of the dynamics must be ensured via other approximations. The dynamics then presents features reminiscent from strong coupling to the environment: e.g. the equilibrium state with an environment at thermal equilibrium is not a Gibbs state of the local system Hamiltonian and the dynamics couples energy coherences and populations. Nonetheless, we build a unified, consistent thermodynamic framework for those dynamics. We focus on two approximations beyond the full secular approximation (the local and the more accurate geometric-arithmetic approximations), and then employ systematic perturbation theory on microscopic expressions for the first and second law of thermodynamics which relate irreversibility to the increase of system-environment correlations. We show that, in spite of the weak system-bath coupling, the system-bath interaction energy participates to the energy balance, and is directly linked to the Lamb-shift Hamiltonian. We compare this microscopic entropy production with the definition based on the contractivity of the reduced system dynamics (Spohn inequality). We show that, unlike for secular master equations, the two entropy production rates differ because of the presence of non-vanishing stationary coherences in the energy eigenbasis. However, in the case of a single thermal bath, the difference is purely transient, and no work can be cyclically extracted from the steady-state despite its non-Gibbs form. Finally, we illustrate our results with simple examples, clarifying and completing the thermodynamic picture of Markovian dynamics in the quantum regime.
\end{abstract}

\maketitle

\section{Introduction}

The thermodynamic description of open quantum systems is mainly based on quantum master equations derived in the weak-coupling and Markovian regime \cite{Breuer_2002,Schlosshauer}. In this framework, irreversibility is commonly characterized through entropy production \cite{Landi_21}. For a system coupled to thermal reservoirs, two notions of entropy production are commonly used. The first one is the microscopic entropy production associated with the build-up of system–environment correlations during the global unitary evolution. It can be expressed as the sum of the entropy variation of the system and the entropy flow from the environments \cite{Esposito_2010}. The second one is the entropy production derived from the Spohn inequality, which quantifies the monotonic relaxation of the reduced density matrix toward the stationary state of the master equation \cite{Spohn_78,Spohn_78b}. 
Those notions are well understood for master equations resulting from the secular approximation \cite{Cohen_98}, which is valid whenever all energy transitions in the system are well separated with respect to their respective linewidth (such that rapidly oscillating cross-terms coupling those transitions can be neglected) \cite{Davies_74,Davies_76,Lindblad_76,Gorini_76}. In this limit, coherence is effectively decoupled from the dissipative dynamics, and the thermodynamics closely resembles classical stochastic thermodynamics. As a consequence, the stationary state of the dynamics is of the Gibbs form with the local system Hamiltonian, and in turn, the two definitions of entropy production coincide, providing
a clear and consistent thermodynamic interpretation of Markovian open-system dynamics \cite{Alicki_1979,alicki2018introductionquantumthermodynamicshistory}. 


However, the secular approximation becomes inaccurate in many physical systems of current interest. This is the case for instance in many-body systems where the spectrum becomes exponentially dense with the number of particles. Other relevant cases are weakly interacting bipartite systems and rapidly driven systems, for which the spectrum typically contains nearly degenerate eigenstates. For those, coherent transitions can evolve on timescales comparable to dissipation. Therefore they can significantly influence the dynamics and must be retained in the reduced description of the system.

The first nonsecular master equation goes back to the Redfield equation \cite{REDFIELD19651} that has been used to predict quantum dissipative processes \cite{LI2008220,Soare2011,Akihito_2009,YANG2002355}. Unfortunately, it suffers from being not completely positive and may therefore predict nonphysical scenarios \cite{Breuer_2002,Dumcke_79,Suarez_92,Pechukas_94}. After that, several nonsecular approaches have been developed to address this regime while solving the positivity issue. A successful one is the coarse-grained master equation (CGME): the evolution of the system is averaged over a small time interval in order to smooth out rapid oscillations that can be responsible for positivity violations \cite{Schaller_08,Majenz_13,cresser2017,Farina_19,Mozgunov2020}.

Another class of nonsecular master equations is based on the partial secular approximation (PSA) \cite{Farina_19,Cattaneo_2019}. These equations can be derived from the CGME \cite{cresser2017,Farina_19,Elouard_2020,Davidovic2020}, directly from the Redfield equation \cite{Jeske_15,Gediminas_18,McGauley_20,Nathan_20,Trushechkin_21}, or through full counting statistics arguments \cite{Potts_2021,Soret_22}. Within the PSA, rapidly oscillating terms are discarded while slowly varying contributions are retained, together with additional assumptions on the environmental spectral density, which is taken to vary slowly around the system’s large Bohr frequencies. Among these approaches, geometric approximations of the decay rates \cite{Gediminas_18,Nathan_20,Davidovic2020,Soret_22} are particularly promising, as they can outperform CGMEs in accuracy while relying only on the Markov approximation.

A related framework corresponds to the local master equation \cite{Mari_12,Wichterich_07,DECORDI2017366,Gonzalez_2017,Cattaneo_2019,Trubilko_2020,Scali2021}, typically used for weakly interacting multipartite systems whose Hamiltonians cannot be easily diagonalized. In this case, the exact eigenoperators are approximated by local operators acting only on the subsystems directly coupled to the environment. However, the term “local master equation” is not used consistently in the literature. Some PSA-based equations are called local because the exact eigenoperators can be summed into clusters that reproduce local operators \cite{Hofer_2017,Elouard_2020,Potts_2021,Trushechkin_21}. In other cases, equations are labeled local without explicitly stating whether secular or PSA approximations are employed. Here we follow the nomenclature of Ref.~\cite{Cattaneo_2019}: a master equation is called local (global) when its eigenoperators act locally (globally) on the system, and we explicitly specify whether the secular approximation or PSA is used. Under this convention, the equations in \cite{Hofer_2017,Elouard_2020,Potts_2021,Trushechkin_21} are local with PSA, whereas geometric master equations (GMEs) are global with PSA.

Although a substantial effort has been devoted for deriving accurate master equations beyond the secular regime and clarifying their dynamical interpretation, their thermodynamic description remains incomplete. The main difficulty is that the nonsecular dynamics, despite following from a weak-coupling expansion, exhibits a behavior reminiscent to open system strongly coupled to reservoirs \cite{Ludovico_14,Esposito_15,Esposito_15b}. First, the dynamics of the energy populations and coherences become interdependent on each other, such that the process cannot anymore be interpreted as a classical Markov chain in a fixed basis. Second, the Lamb shift Hamiltonian and the dissipative part of the dynamics do not commute anymore with the bare system Hamiltonian. Moreover, the stationary state of the dynamics is not anymore a Gibbs state of the system Hamiltonian, even when the reservoir is at thermal equilibrium \cite{Thingna_12,Lobjeko_24}. 

Those properties have made difficult a consistent thermodynamic description of this regime to emerge, which was frequently categorized as being thermodynamically inconsistent. The first reported violation of the second law appeared in Ref.~\cite{Levy_2014} for a local master equation describing a weakly interacting bipartite system coupled to thermal baths. However, this violation was only apparent: the considered equation was effectively secular, where the second law still holds, and the inconsistency arose from using the local approximation beyond its validity range. A consistent perturbative treatment resolves the issue \cite{Trushechkin_2016}. Nevertheless, this example highlights the need for care when formulating the first and second laws for a given dynamics.

Subsequent works proposed solutions for thermodynamic inconsistencies in local nonsecular master equations (LNMEs), where genuinely quantum contributions to entropy production and energy may arise \cite{Mitchison_2018,Elouard_2020,Soret_25}. These approaches can be divided into two categories. The first aims to restore consistency for a given master equation. Different proposals exist, but no consensus has emerged. For example, Ref.~\cite{Potts_2021} introduces a bookkeeping Hamiltonian so that entropy production matches the resolution of the master equation, whereas Refs. ~\cite{DeChiara_2018,Hewgill_21} reinterpret the interaction energy between subsystems (usually responsible for violations) as a work contribution.

The second category identifies general dynamical properties ensuring thermodynamic consistency and then deduces the form of the master equation satisfying those criteria. Ref.~\cite{Dann_21}, for instance, imposes strict energy conservation (that is, does not allow the interaction energy to participate into the energy balance) and concludes that, for a system coupled to a single bath, only the global secular master equation is admissible. Very recently, a proposed unification of Markovian quantum thermodynamics identifies thermodynamically admissible dynamics with Markovian thermal operations (i.e. evolutions preserving the Gibbs state of the local system Hamiltonian) \cite{luo2026trinitymarkovianquantumthermodynamics}. However, those requirements are arguably overly restrictive, since LNMEs can also be thermodynamically consistent even without necessarily satisfying these criteria \cite{Potts_2021,DeChiara_2018,Soret_22}. For instance, strict energy conservation is stronger than the first-law requirement of conserving the total energy (including the coupling energy).

At the same time, while the Spohn inequality still guarantees positivity of an entropy-production-like quantity, its relation to the microscopic entropy production associated with system–environment correlations and its energetic interpretation have not been fully clarified. Discrepancy between both have been reported in few cases such as in the Bloch optical equation \cite{Elouard_2020}, or for the CGME \cite{Schaller_20}, while a detailed analysis is still lacking.
More generally, the thermodynamic role played by the coherent corrections induced by nonsecular transitions and potentially the Lamb-shift terms remains unclear.

In this work, we develop a consistent thermodynamic description of nonsecular master equations at the average level. Although we focus on the GMEs and local master equations as prototypical classes of nonsecular dynamics, the thermodynamic construction is not specific to them. Indeed, it follows from the underlying weak-coupling, coarse-grained system–bath dynamics and applies, with the corresponding perturbative order maintained consistently, to the CGME and to other nonsecular Markovian master equations derived within the same framework. 
Our treatment identifies the thermodynamic counterparts to the phenomena specific to the nonsecular dynamics, while showing they comply with the laws of thermodynamics when the expressions for the work and the heat flows in term of the system dynamics are rigorously derived, with the same order of magnitude as the dynamics.

First, we analyze energy conservation between the system and the bath, and show that the interaction energy and the Lamb-shift energy both play a significant role in the balance. More precisely, the coupling energy flow splits into two equal halves which each contribute to the energy flows received by the system and the bath, respectively, and happen to be equal to the flow of the energy stored in the Lamb-shift Hamiltonian.

We then formulate the second law by deriving the microscopic entropy productions associated with the increase of system-bath correlations \cite{Esposito_2010}, and  based on the Spohn inequality, respectively. We show that in the nonsecular case, the two entropy productions differ and relate this discrepancy to the non-commuting Lamb-shift and the non-Gibbs stationary state.
We further show that, despite these aspects resemble nonequilibrium steady states (NESS) \cite{yukawa2001,Horowitz_14,Manzano_18}, they are still thermodynamically compatible with the contact with a single thermal reservoir. Indeed, we compute the work flow and show that when it is nonzero, it is only transient and no net work may be cyclically extracted from the single reservoir. On the way, we also recover a consistent thermodynamic description for LNMEs as a limiting case of our analysis of the GME. The resulting description agrees with the approach of Potts et al. \cite{Potts_2021}, allowing to complete the thermodynamic picture of nonsecular dynamics, from global to local.

The manuscript is organized as follow: in Sec.~\ref{sec.II} we describe the dynamics of a quantum system weakly coupled to a thermal bath by starting from the CGME, and we end with the geometric arithmetic master equation (GAME) by following Ref.~\cite{Davidovic2020}. We also show how LNMEs can be derived from GAME by taking the small Bohr frequencies of the system to zero. In Sec.~\ref{sec.III} we present a consistent thermodynamics description for GAME, where in Sec.~\ref{sec.IIIA} we write the conservation of energy. In Sec.~\ref{sec.IIIB} we derive the 2nd law from the microscopic entropy production and compare to the entropy from the Spohn inequality in Sec.~\ref{sec.IIIC}. We then extend our results to the local regime in Sec.~\ref{sec.IIID}, while Sec.~\ref{sec.IIIE} focuses on the 1st law. In Sec.~\ref{sec.IIIG} we extend our results to the case of multiple baths and then illustrate our findings with simple examples in Sec.~\ref{sec.IV}. Finally we conclude in Sec.~\ref{sec.V}.

\section{Dynamics of the system}\label{sec.II}
\subsection{Coarse-grained master equation}
We consider an open quantum system described by the total Hamiltonian
\begin{equation}
    H=H_{S}+H_{B}+H_{I},
\end{equation}
where $H_{S}$ is the Hamiltonian of the system, $H_{B}$ is the Hamiltonian of the bath and $H_{I}=A\otimes B$ the coupling Hamiltonian with $A$ acting on the system and $B$ acting on the bath. All three parts are assumed to be independent of time.
The environment is assumed to be at thermal equilibrium, such that its initial state is given by the Gibbs density matrix $\rho_{B}=Z_{B}^{-1}\exp(-\beta H_{B})$ (with $Z_{B}$ the partition function and $\beta$ the inverse temperature). 
We describe the dynamics of the system within the Born-Markov approximation by deriving a coarse-grained master equation (CGME) in the interaction picture (see App.~\ref{appA} for the derivation or \cite{Cohen_98} for a thorough treatment)
\begin{equation}\label{CGME}
    \frac{\Delta\tilde{\rho}_{S}}{\Delta t}=-i[\tilde{H}^{\Delta t}_{LS},\tilde{\rho}_{S}]+\tilde{D}^{\Delta t}(\tilde{\rho}_{S}),
\end{equation}
where the tilde denotes operators written in the interaction-picture, and we took $\hbar=1$. The master equation contains two parts: the first one is unitary and is related to a coarse-grained Lamb-shift given by 

\begin{equation}
    \tilde{H}^{\Delta t}_{LS}= \sum_{\omega,\omega'} f_{\Delta t}(\omega,\omega')S(\omega,\omega')e^{i(\omega'-\omega)t}A^{\dagger}(\omega')A(\omega)
\end{equation}
where $f_{\Delta t}(\omega,\omega')=e^{i(\omega'-\omega)\Delta t/2}\text{sinc}((\omega'-\omega)\Delta t/2)$, $A(\omega)$ is the eigenoperator of $H_S$ at the eigenfrequency $\omega$ i.e $[H_{S},A(\omega)]=-\omega A(\omega)$ and $S(\omega,\omega')$ is the antisymmetric part of the half Fourier transform of the bath correlation function matrix:

\begin{equation}\label{SLS}
    S(\omega,\omega')=\frac{\Gamma(\omega)-\Gamma^{*}(\omega')}{2i}
\end{equation}
where

\begin{equation}
\begin{split}
    \Gamma(\omega)&=\int_{0}^{\infty}e^{i\omega\tau}g(\tau)d\tau,\\
    g(\tau)&=\Tr_{B}\left(\tilde{B}(\tau)B\rho_{B}\right).    
\end{split}
\label{SDF}
\end{equation}

The second term of the master equation is non-unitary and is given by the coarse-grained dissipator 

\begin{equation}\label{D}
\begin{split}
    \tilde{D}^{\Delta t}(\tilde{\rho}_{S})=&\sum_{\omega,\omega'}f_{\Delta t}(\omega,\omega')\gamma(\omega,\omega')e^{i(\omega'-\omega)t}\left(A(\omega)\tilde{\rho}_{S}A^{\dagger}(\omega')\right.\\
    &\left.-\frac{1}{2}\left\{A^{\dagger}(\omega')A(\omega),\tilde{\rho}_{S}\right\}\right),
\end{split}
\end{equation}
where
\begin{equation}\label{gamma2}
    \gamma(\omega,\omega')=\Gamma(\omega)+\Gamma^{*}(\omega'),
\end{equation}
is the symmetric part of the half Fourier transform of the bath correlation function matrix, and corresponds to the decay rate.

In this approach, the evolution of the system is coarse grained in time in the sense that the derivative of the density matrix is approximated by its average over a coarse-graining time $\Delta t$ \cite{Cohen_98}
\begin{equation}
    \frac{\Delta\tilde\rho_{S}}{\Delta t}=\frac{1}{\Delta t}\int_{t}^{t+\Delta t}\frac{d\tilde\rho_{S}}{dt'}dt'.
\end{equation}
The coarse-graining time is chosen to smooth out rapid oscillations of the instantaneous state of the system. Within the Born-Markov approximation we can choose it to be both longer than $\tau_{B}$, the typical time at which the bath correlation functions decay, and shorter than the evolution timescale of the system in the interaction picture, $\tau_{R}$. The latter corresponds to the slow dissipation induced by the weak coupling to the thermal bath and can be written as $\tau_{R}=\gamma^{-1}$, with $\gamma\sim\lambda^{2}\tau_{B}$ the typical relaxation rate of the dynamics, and $\lambda$ the coupling strength between the system and the bath. The coarse-grained description is accurate whenever
\begin{equation}\label{condition}
    \tau_{B}\ll\Delta t\ll \gamma^{-1}.
\end{equation}

Assuming that Eq.~\eqref{CGME} captures well the evolution of the system in the interaction picture, we can then write the corresponding master equation in the Schr\"odinger picture by 
\begin{equation}
    \frac{d\rho_{S}}{dt}\approx-i[H_{S}+H^{\Delta t}_{LS},\rho_{S}]+D^{\Delta t}(\rho_{S}).
\end{equation}

\begin{figure*}[!t]
  \centering
  \includegraphics[width=0.8\textwidth]{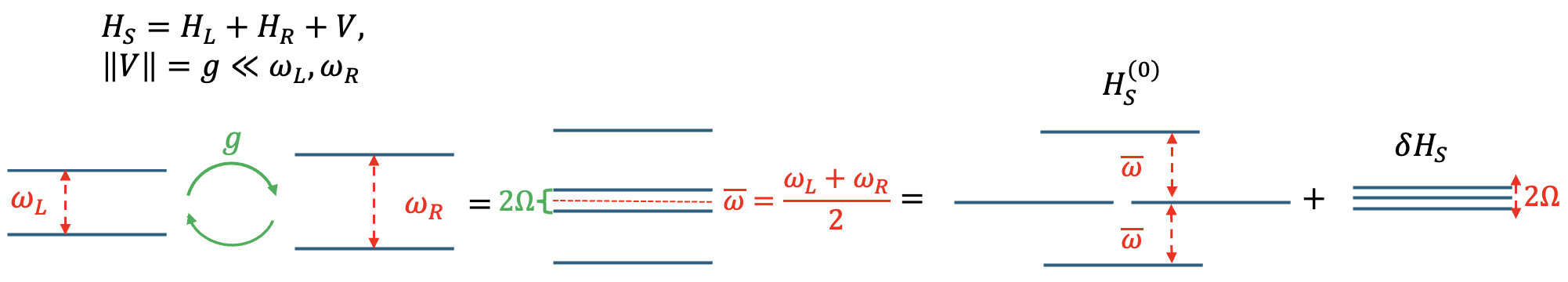}
  \caption{An illustrative example of a quantum system where the PSA is relevant. Here we have two qubits that interact such that the coupling strength is very small compared to the frequencies. The spectrum will thus contain two close eigenstates such that we can identify $H^{(0)}_{S}$ with the dominant Bohr frequencies $\pm\overline{\omega}$ and the perturbation $\delta H_{S}$ with Bohr frequencies $\pm2\Omega$ satisfying $\Omega\ll\overline{\omega}$.}
  \label{Fig2B1}
\end{figure*}

The case $\Delta t=0$ (no coarse-graining) corresponds to the Bloch-Redfield equation, which is known to break positivity of the dynamics \cite{Dumcke_79,Pechukas_94}. In contrast, for $\Delta t>0$, the decay rates appearing in dissipator Eq.~\eqref{D} are rescaled by a factor $f_{\Delta t}(\omega,\omega')$. As a consequence, one may find an appropriate choice of the coarse-graining time, depending on the parameters of the system and the bath, ensuring a complete positive dynamics \cite{Schaller_08,Farina_19}. This is the main advantage of the CGME. In particular, the (full) secular approximation corresponds to the case where one assumes that $\Delta t(\omega-\omega')\gg 1$, $\forall$ $\omega\neq\omega'$, then $f_{\Delta t}(\omega,\omega')\rightarrow\delta_{\omega,\omega'}$ and one recovers the secular Lindblad equation
\begin{equation}
\begin{split}
    \frac{d\rho_{S}}{dt}&=-i[H_{S}+H_{LS},\rho_{S}]+D^{L}(\rho_{S}),\\
    H_{LS}&=\sum_{\omega} S(\omega)A^{\dagger}(\omega)A(\omega),\\
    D^{L}(\rho_{S})&=\sum_{\omega}\gamma(\omega)\left(A(\omega)\rho_{S}A^{\dagger}(\omega)\right.\\
    &\left.-\frac{1}{2}\left\{A^{\dagger}(\omega)A(\omega),\rho_{S}\right\}\right).
\end{split}
\end{equation}
In the secular Lindblad equation, the Lamb-shift commutes with the system Hamiltonian and therefore reduces to a shift of its eigenenergies while the dissipator implies well-defined transitions between the eigenstates of the system, similarly to a classical stochastic process.

\subsection{From the coarse-grained to the geometric-arithmetic master equation}

Here, we want to study the dynamics of a system beyond the secular Lindblad equation, i.e. when the secular approximation is not valid anymore. One possibility is to consider the CGME with a given choice of $\Delta t$ treated as a parameter. However, finding the value for which complete positivity of the dynamics is preserved without deteriorating too much the accuracy of the master equation is only tractable analytically for simple systems and can become rapidly demanding for large systems. Moreover, directly employing the CGME necessitates to accept system dynamics, and consequently thermodynamics observables, that explicitly depend on the value of $\Delta t$, even in the stationary limit, which seems unphysical. Also, let us mention that the thermodynamics of CGMEs has been studied with a focus on the entropy production, where the system-bath coupling is assumed to be turned on for a time interval of duration $\Delta t$ \cite{Schaller_20}.

In order to overcome this issue, we need to use a partial secular approximation (PSA), that will strike balance between dynamical accuracy and positivity preservation of the state of the system, and remove the dependency on $\Delta t$. For that we need to specify more how the spectrum of the system behaves. Here, we assume that we can identify a set of dominant frequencies $\{\overline{\omega}\}$ such that the Bohr frequencies of the system $\omega$ can be gathered into clusters $\mathcal{L}(\overline\omega)$ in the following way

\begin{equation}
\begin{split}
    |\omega'-\omega|\Delta t&\ll1 \; \text{for} \; \omega,\omega'\in \mathcal{L}(\overline\omega)\\
    |\omega'-\omega|\Delta t&\gg1 \; \text{for} \; \omega\in \mathcal{L}(\overline{\omega}_{i}),\omega'\in\mathcal{L}(\overline{\omega}_{j}) \; \text{and}\;  \overline{\omega}_{i}\neq\overline{\omega}_{j}.
\end{split}
\end{equation}
In other words, the spectrum of the system can be split into sets of nearly-degenerate frequencies each around one of the values $\overline{\omega}$. Such a situation occurs in many cases. One example is a system weakly perturbed by a quasi-resonant external drive \cite{Elouard_2020,Prasad_26}. More commonly in thermodynamics and quantum transport, it also typically arises for a system made of $N$ subsystems that weakly interact between each other, i.e the Hamiltonian is of the form $H_{S}=\sum_{n}H_{n}+V$ with $H_{n}$ describing a subsystem, and $V$ the interaction term such that $\lVert V\rVert\ll \lVert H_{n}\rVert$. After diagonalizing it, one can identify a dominant term and a perturbation term $H_{S}=H_{S}^{(0)}+\delta H_{S}$ with $[\delta H_S,H_{S}^{(0)}]=0$. In that case, the dominant frequencies $\overline\omega$ correspond to the Bohr frequencies of $H_{S}^{(0)}$ \cite{Trushechkin_21} (see Fig.~\ref{Fig2B1} for an illustrative example).

At this point, a common version of the PSA consists in replacing $f_{\Delta t}(\omega,\omega')$ by $1$ when $\omega,\omega'$ belong to the same cluster, and $0$ otherwise. However, as it will be shown later, the thermodynamic analysis stemming from this approach has a limited accuracy. We therefore instead follow the more accurate procedure proposed by Davidovic, leading to the so-called geometric-arithmetic master equation (GAME) \cite{Davidovic2020}. The core idea consists in keeping only the dominant contribution of the decay rates, which is symmetric and preserves the positivity of the system dynamics, while removing the part that is responsible for breaking the positivity and that is actually negligible for a sufficiently large value of $\Delta t$. More precisely, let us introduce the real and imaginary parts of $\Gamma(\omega)$  
\begin{equation}\label{BathFourier}
\begin{split}
    \Gamma(\omega)&=\frac{\gamma(\omega)}{2}+iS(\omega),\\
    \gamma(\omega)&=\int_{-\infty}^{\infty}e^{i\omega\tau}g(\tau),\\
    S(\omega)&=\frac{1}{2\pi}\mathcal{P}\int_{0}^{\infty}\frac{\gamma(\omega')}{\omega-\omega'}d\omega',
\end{split}
\end{equation}
where $\mathcal{P}$ is the Cauchy principal value. Using Eq.~\eqref{gamma2}, we can then rewrite the decay rates using the identity
\begin{equation}
\begin{split}\label{gamma}
    f_{\Delta t}(\omega,\omega')\gamma(\omega,\omega')=f_{\Delta t}(\omega,\omega')\left(\sqrt{\gamma(\omega)\gamma(\omega')}\right.\\
    \left.+\frac{\left(\sqrt{\gamma(\omega)}-\sqrt{\gamma(\omega')}\right)^{2}}{2}+i\left(S(\omega)-S(\omega')\right)\right).
\end{split}
\end{equation}

For two Bohr frequencies $\omega,\omega'$ that do not belong to the same cluster $\mathcal{L}(\overline{\omega})$ we have $|\omega-\omega'|\Delta t\gg1$ and therefore the decay rate vanishes due to the sinc function contained in $f_{\Delta t}(\omega,\omega')$. Conversely, when the two frequencies are in the same cluster, $f_{\Delta t}(\omega,\omega')\rightarrow 1$ and the symmetric term $\sqrt{\gamma(\omega)\gamma(\omega')}$ dominates the two last terms of the right-hand side of Eq.~\eqref{gamma}. Indeed, for a coarse-graining time such that $|\overline{\omega}|\Delta t\gg1$ and $|\omega-\omega'|\Delta t\ll  1$, we can evaluate those second and third terms from a Taylor expansion around $\overline{\omega}$, yielding at first order
\begin{equation}
\begin{split}
    &\left|f_{\Delta t}(\omega,\omega')\left(\frac{\left(\sqrt{\gamma(\omega)}-\sqrt{\gamma(\omega')}\right)^{2}}{2}+i\left(S(\omega)-S(\omega')\right)\right)\right|\\
    &\approx |\omega-\omega'||S'(\overline{\omega})|+\mathcal{O}(|\omega-\omega'|^{2})\ll\frac{|S'(\overline\omega)|}{\Delta t}.
\end{split}
\end{equation}
The last inequality implies that those two terms decay at least as $1/\Delta t$ and can be neglected for sufficiently large $\Delta t$. Note that one does not necessarily needs to assume that the frequencies $\omega,\omega'$ are close to each other in order to neglect these contributions. Indeed, it is sufficient to assume that $\Delta t$ is large enough with respect to inverse of the typical frequency scale over which the spectral density varies, which can be formally related to the Markov approximation (see Ref.~\cite{Nathan_20} where the validity of the geometric approximation stems solely from Markovianity). Moreover, Davidovic numerically showed that taking $\Delta t>\tau_{B}$ is enough for these terms to be neglected, such that it requires no additional assumption with respect to condition Eq.~\eqref{condition}. In summary, the geometric-arithmetic approximation allows us to write the decay rates as follows
\begin{equation}\label{geom_approx}
    f_{\Delta t}(\omega,\omega')\gamma(\omega,\omega')\approx\begin{cases}\sqrt{\gamma(\omega)\gamma(\omega')}\; \text{if} \; \omega,\omega'\in\mathcal{L}(\overline{\omega}).\\
    0\;\text{if}\; \text{not}.
      \end{cases}
\end{equation}
Furthermore, the Lamb-shift coefficients are approximated as
\begin{equation}
    f_{\Delta t}(\omega,\omega')S(\omega,\omega')\approx\begin{cases}S(\omega,\omega')\; \text{if} \; \omega,\omega'\in\mathcal{L}(\overline{\omega}).\\
    0\;\text{if}\; \text{not},
      \end{cases}
\end{equation}
and we finally obtain the GAME describing the dynamics of the system
\begin{equation}\label{GAME}
    \begin{split}
        \frac{d\rho_{S}}{dt}&\approx-i\left[H_{S}+H_{LS},\rho_{S}\right]+D^{G}(\rho_{S}),\\
        H_{LS}&=\sum_{\overline{\omega}}\sum_{\omega,\omega'\in\mathcal{L}(\overline{\omega})}S(\omega,\omega')A^{\dagger}(\omega')A(\omega),\\
        D^{G}(\rho_{S})&=\sum_{\overline{\omega}}\sum_{\omega,\omega'\in\mathcal{L}(\overline{\omega})}\sqrt{\gamma(\omega)\gamma(\omega')}\left(A(\omega)\rho_{S} A^{\dagger}(\omega')\right.\\
        &-\left.\frac{1}{2}\left\{A^{\dagger}(\omega')A(\omega),\rho_{S}\right\}\right).
    \end{split}
\end{equation}
It has been shown that the GAME has a better accuracy of the dynamics than the CGME, even taken with an optimal coarse-graining time \cite{Davidovic2020}. We note that, contrary to the secular Lindblad equation, the Lamb-shift term of the GAME contains off-diagonal terms and does not necessarily commute with $H_{S}$. Moreover, the dissipator can be simply rewritten in a diagonal form as
\begin{equation}
    D^{G}(\rho_{S})=\sum_{\overline\omega}A_{\overline{\omega}}\rho_{S}A^{\dagger}_{\overline{\omega}}-\frac{1}{2}\{A^{\dagger}_{\overline{\omega}}A_{\overline{\omega}},\rho_{S}\},
\end{equation}
with
\begin{equation}
    A_{\overline{\omega}}=\sum_{\omega\in\mathcal{L}(\overline{\omega})}\sqrt{\gamma(\omega)}A(\omega).
\end{equation}
Written this way, we clearly see that the dynamics involves jumps that are superpositions of eigenoperators belonging to the same cluster $\mathcal{L}(\overline{\omega})$ and therefore do not correspond to well-defined transitions in the eigenbasis of the system. Indeed, the jump operators $A_{\overline{\omega}}$ are not eigenoperators of $H_{S}$ in general (this is why we put $\overline{\omega}$ as a subscript). Therefore, the dynamics cannot be reduced to a stochastic classical process.

\subsection{The local nonsecular master equation as a limiting case of the GAME}\label{sec.IIC}

From the GAME we can recover the local nonsecular master equation (LNME) for which different consistent thermodynamic descriptions have been proposed \cite{Potts_2021,DeChiara_2018,Hewgill_21}. Indeed, assuming that the Bohr frequencies belonging in the same set $\mathcal{L}(\overline{\omega})$ are close enough to $\overline\omega$ such that the decay rates and the Lamb-shift coefficients can be approximated at $\overline{\omega}$ i.e $\gamma(\omega)\approx\gamma(\overline{\omega})$ and $S(\omega,\omega')\approx S(\overline\omega,\overline{\omega})=S(\overline{\omega})$ for $\omega\in\mathcal{L}(\overline\omega)$ we get

\begin{equation}\label{LNME}
    \begin{split}
        \frac{d\rho_{S}}{dt}&\approx-i\left[H_{S}+H^{(0)}_{LS},\rho_{S}\right]+D^{(0)}(\rho_{S}),\\
        H^{(0)}_{LS}&=\sum_{\overline{\omega}}S(\overline\omega)\left(A^{(0)}_{\overline{\omega}}\right)^{\dagger}A^{(0)}_{\overline{\omega}},\\
        D^{(0)}(\rho_{S})&=\sum_{\overline{\omega}}\gamma(\overline{\omega})\left(A^{(0)}_{\overline{\omega}}\rho_{S} \left(A^{(0)}_{\overline{\omega}}\right)^{\dagger}\right.\\
        &-\left.\frac{1}{2}\left\{\left(A^{(0)}_{\overline{\omega}}\right)^{\dagger}A_{\overline{\omega}}^{(0)},\rho_{S}\right\}\right),
    \end{split}
\end{equation}
with $A^{(0)}_{\overline\omega}=\sum_{\omega\in\mathcal{L}(\overline{\omega})}A(\omega)$. This equation is also called the unified master equation by Trushechkin \cite{Trushechkin_21}. In this regime the bath can not discriminate transitions induced by eigenopertors $A(\omega)$ that belong to the same $\mathcal{L}(\overline\omega)$. Also we see that contrary to the GAME, we can clearly identify jumps $A_{\overline{\omega}}^{(0)}$ that occurs at a well-defined rate $\gamma(\overline{\omega})$. Here the term ``local'' refers to the fact that, in the case where the system Hamiltonian is of the form $H_{S}=\sum_{n}H_{n}+V=H_{S}^{(0)}+\delta H_{S}$, and the system-bath coupling involves only a subpart of the subsystems $n$, the jump operators $A_{\overline{\omega}}^{(0)}$ actually act only on that subpart of the system. In other word, the dissipator takes the same form as if there were no coupling to the rest of the system. This is in contrast with the secular Lindblad equation, in this case referred to as the ``global'' master equation, which is generated by jump operators $A(\omega)$ acting on all subsystems. Because the GAME has a higher accuracy than the LNME, its thermodynamic description will also contain more information, as we detail below.

\section{Thermodynamics}\label{sec.III}
\subsection{Conservation of energy}\label{sec.IIIA}
In this section, we focus on the average values of the thermodynamic observables. The GAME is used below to write the thermodynamic quantities explicitly. However all the formula follow from the coarse-grained system–bath expansion before the specific geometric-arithmetic factorization is imposed. These relations therefore carry over to other nonsecular Markovian master equations, provided that the dynamical, bath-energy, and interaction-energy contributions are truncated consistently at the same perturbative order.
Here we present the results that we derived for the energy conservation and how the different contributions are related to each other, but the technical details are provided in the appendix \ref{appB}.

One remarkable feature of the secular Lindblad equation is that the energy variation of the system is exactly compensated by the energy variation of the bath, despite the presence of the coupling Hamiltonian $H_I$ (which does not necessarily commutes with $H_S+H_B$). This property has been referred to as ``strict energy conservation'' in \cite{Soret_22}, and can be seen as an emergent consequence of the full secular approximation which implies a simplified thermodynamic description, in particular at the stochastic level. However, this property does not hold anymore for GAME, or other nonsecular master equations, implying that the interaction energy appears in the energy balance:
%
\begin{equation}\label{eq:en_bal}
    \frac{dE_{S}}{dt}+\frac{dE_{B}}{dt}+\frac{dE_{I}}{dt}=0.
\end{equation}
Above, the first term corresponds to the energy flow in the system and is simply obtained from the GAME as
\begin{equation}
\begin{split}
    \frac{dE_{S}}{dt}&=\Tr\left(H_{S}\frac{d\rho_{S}}{dt}\right)\\
    &=-i\Tr\left(H_{S}\left[H_{LS},\rho_{S}\right]\right)+\Tr\left(H_{S}D^{G}(\rho_{S})\right).
\end{split}
\end{equation}
We see that unlike the secular case, the system has an energy contribution coming from the Lamb-shift as the latter does not necessarily commute with $H_{S}$. In other words, the energy change of the system is not purely dissipative. The second term is the energy flow in the bath. We calculate it from the coarse-grained evolution of $\rho_{SB}$
\begin{equation}\label{eq:dEbGAME}
    \frac{dE_{B}}{dt}\approx\frac{\Tr\left(H_{B}(\tilde\rho_{SB}(t+\Delta t)-\tilde\rho_{SB}(t))\right)}{\Delta t}.
\end{equation}
As shown in App.~\ref{appB}, the same approximations involved in the derivation of the master equation allow us to express the bath energy flow only in terms of the system observables

\begin{equation}\label{dEB2}
    \frac{dE_{B}}{dt}=-i\Tr_{S}\left(H_{S}\left[H_{LS},\rho_{S}\right]\right)-\Tr_{S}\left(H_{S}D^{G}(\rho_{S})\right).
\end{equation}
We see that, like the system, the bath energy flow contains a contribution coming from the Lamb-shift (with same sign as that appearing with in the expression of the system flow) and one coming from the dissipator (of opposite sign with respect to the system). This is also quite different from the secular case where the same quantity is only related to the dissipator and is exactly equal to the opposite of the energy flow in the system, allowing to conclude with no ambiguity that the system only exchanges heat with the bath. Similarly to the bath energy flow, the coupling energy flow is obtained from the coarse-grained evolution of the system and the bath via:
\begin{equation}\label{eq:EIflow1}
    \frac{dE_{I}}{dt}\approx\frac{\Tr\left(\tilde H_{I}(t+\Delta t)\tilde\rho_{SB}(t+\Delta t)-\tilde H_{I}(t)\tilde\rho_{SB}(t)\right)}{\Delta t}.
\end{equation}
Combining the previous equations, we can simply deduce that this flow is related to the Lamb-shift energy
\begin{equation}\label{eq:EIflow2}
    \frac{dE_{I}}{dt}=2i\Tr_{S}\left(H_{S}\left[H_{LS},\rho_{S}\right]\right).
\end{equation}
For completeness, we also derived Eq.~\eqref{eq:EIflow2} from Eq.~\eqref{eq:EIflow1} and the approximations leading to the master equations, see  App.~\ref{appB}. We see that, surprisingly, the coupling energy flow is equally split between the contributions received by the system and by the bath. Moreover, the expression $-i\Tr\left(H_{S}\left[H_{LS},\rho_{S}\right]\right)$ allows to relate it to the energy variation due to the Hamiltonian evolution induced by the Lamb-shift. This explicit connection between the Lamb-shift and the coupling energy is the main results of this section. It outlines that one needs to be careful when neglecting the Lamb-shift in deriving a quantum master equation, as it here play a fundamental role from an energetic and entropic point of view (see also \cite{Schaller_20}).

Finally, we finish this discussion by showing what happens when moving from the GAME to the LNME (Eq.~\eqref{LNME}). As we showed in the previous section, the LNME is a zero order approximation of the GAME around the dominant frequencies. Deriving the corresponding energy flows, while keeping a consistent level of approximation, requires a careful analysis of the order of magnitudes. This is even more the case since the master equation is already an approximated equation of the real dynamics of the system, already involving an expansion. Indeed, within the Born-Markov approximation, it corresponds to a first order expansion of the real dynamics with respect to the typical relaxation rate $\gamma$  (up to the second order in terms of the interaction Hamiltonian). To do so, we denote with a superscript $(k)$ the $k$th order expansion of the dynamics with respect to $\gamma$, and with a superscript $(k,j)$ the order $k$ with respect to $\gamma$, and $j$ with respect to $\delta H_S$ (i.e. with respect to the magnitude of the energy differences inside the frequency clusters). The GAME then takes the form: 
\begin{equation}\label{me_expansion}
\begin{split}
    \frac{d\rho_{S}}{dt}&\approx L^{(0)}(\rho_{S})+\gamma L^{(1)}(\rho_{S}),\\
    L^{(0)}(\rho_{S})&=-i\left[H_{S},\rho_{S}\right],\\
    L^{(1)}(\rho_{S})&=-i\left[H_{LS},\rho_{S}\right]+D^{G}(\rho_{S}),
\end{split}
\end{equation}
while the LNME can be written as
\begin{equation}
\begin{split}
    \frac{d\rho_{S}}{dt}&\approx L^{(0)}(\rho_{S})+\gamma L^{(1,0)}(\rho_{S}),\\
    L^{(1,0)}(\rho_{S})&=-i\left[H^{(0)}_{LS},\rho_{S}\right]+D^{(0)}(\rho_{S}).
\end{split}
\end{equation}
Having in mind that $H_{S}=H_{S}^{(0)}+\delta H_{S}$, the system energy flow of the system as evaluated by the LNME becomes rigorously
\begin{equation}
\begin{split}
    \frac{dE_{S}}{dt}&=\gamma\Tr_{S}(H_{S}L^{1}(\rho_{S}))\\
    &\approx\gamma\Tr_{S}\left(H_{S}^{(0)}L^{(1,0)}(\rho_{S})\right)\\
    &=\gamma\Tr_{S}(H_{S}^{(0)}D^{(0)}(\rho_{S}))\\
    &\equiv\frac{dE_{S}^{(0)}}{dt}.\label{eq:dES_loc}
\end{split}
\end{equation}
We got rid of the term coming from $\delta H_{S}$ when going to the second line since it has the same order of magnitude than contributions coming from the higher order terms of the expansion of $L^{(1)}$ (i.e $L^{(1,j)}$ with $j\geq1$), not present in the LNME. The term associated with the Lamb-shift vanishes since in the LNME regime, $H_{LS}$ commutes with $H_{S}^{(0)}$. As a consequence, we recover strict energy conservation 
\begin{equation}\label{eq:en_bal_loc}
    \dfrac{dE_S^{(0)}}{dt}=-\dfrac{dE_B^{(0)}}{dt}
\end{equation}

This feature allows to give an energetic description of the LNME analogous to that of the secular Lindblad equation. 
This is explained by the fact that both share similar dynamical features even though they do not have the same regime of validity. The secular Lindblad equation describes a classical stochastic process where transition occurs between the eigenstates of $H_{S}$ with rates $\gamma(\omega)$. The LNME also features a classical stochastic process, but the transitions are occuring between the clusters $\mathcal{L}(\overline{\omega})$, that can be related to the eigenstates of $H_{S}^{(0)}$. In other words the LNME describes a regime that is fully secular with respect to frequencies of $H_{S}^{(0)}$, ignoring all the contributions from next orders. This can correspond to two different situations: the first one is when contribution due to $\delta H_{S}$ are indeed negligible with respect to the properties of the bath, as we already mentioned above \cite{Potts_2021,Trushechkin_21,Elouard_2020}. The second situation correspond to a case where we track the dynamics on a timescale over which only the frequencies of $H_{S}^{(0)}$ have been resolved, but not yet the frequencies of the order of $\delta H_{S}$ \cite{Winczewski_24}.      

Our reasoning is in agreement with the thermodynamics framework proposed by Potts et al. \cite{Potts_2021}. The authors introduced a thermodynamic bookkeeping Hamiltonian $H_{TD}$ that ensures that the resolution of the thermodynamic observables are of the same order of magnitude as that of the master equation. This bookkeeping Hamiltonian corresponds here to the zero order Hamiltonian $H^{(0)}_{S}$. At the same time, we can conclude that the approach developed in Refs.~\cite{DeChiara_2018,Hewgill_21}, can not be valid when the environment is a bath a thermal equilibrium. In this work, the authors use the rationale that the nonsecular master equation can also be generated by a collision microscopic model to propose to treat the energy coming from the interaction $V$ as a work cost necessary to maintain the violation of the detailed balance condition with respect to the Bohr frequencies of $H_{S}$ appearing in the LNME.

Our analysis shows that, however, when the nonsecular master equation arises from a bath at equilibrium rather than a collision model, this energy term violates the conservation of energy (in contrast with the case of a collision model, where it signals the existence of a work source switching on and off the coupling between the system and the collision units). As the nonsecular contribution to the energy balance are beyond the resolution of the LNME, a finer dynamical description (like GAME) is necessary to provide a rigorous interpretation of this term, as we do below.

\subsection{Entropy production and second law}\label{sec.IIIB}

Here we analyze the average entropy production for the GAME and show how it differs from the secular Lindblad equation. 
In the context of quantum master equations, two main formalisms have been used to quantify entropy production. On one hand, entropy production can be defined from the system-bath and internal bath correlations induced by the unitary dynamics of the total system \cite{Esposito_2010}. For a bath initially in a thermal equilibrium state at temperature $T$, this leads to define:
\be\label{d:sigmatot}
 \sigma_\text{tot} = \Delta S + \beta\Delta E_B = S\left(\rho_{SB}(t)\Big\|\rho_S(t)\otimes\rho_B^\text{eq}\right) \geq 0,
\ee
as the entropy production between times $0$ (where the system and the bath are uncorrelated) and $t$, where the energy variation of the bath is then identified with heat $\Delta E_B$. 
This general picture has the advantage of being valid for any system and any coupling strength, provided the initial state is factorized between the system and environment spaces. Moreover, it can be extended to any initial environment state, allowing for a general accounting of the work exchanged with the environment when the latter starts or ends in a nonthermal state \cite{Elouard_23}. This approach has the disadvantage of defining heat and entropy production in terms of the environment state. However, in the regime of validity of the master equations considered in this paper, an expression in terms of the sole system observables can be obtained for the entropy production rate $\dot\sigma_\text{tot}$. This is achieved by applying Eq.~\eqref{d:sigmatot} for a time-interval equal to the coarse-graining time $\Delta t$, and then performing on the bath energy flow $\Delta E_B/\Delta t \simeq \dot E_B$ the same approximations as invoked to derive the master equation from the microscopic model \cite{Elouard_2020} (see also Section \ref{sec.IIIA})  
\begin{equation}
    \dot\sigma_\text{tot}(t)=\frac{dS}{dt}+\beta\frac{dE_{B}}{dt}.
\label{total_entropy}
\end{equation}

The heat is thus related to the bath energy change, and the obtained expression then depends on the specific master equation. On the other hand, another strategy consists in defining entropy production from the reduced evolution of the system only, without referring to the underneath specific microscopic model for the environment. In that case, irreversibility (in the sense of time arrow) is associated to the contracting property of CPTP maps, i.e. the rate of the monotonous convergence towards the steady state(s), as quantified by the Spohn inequality \cite{Spohn_78,Spohn_78b}. In the case of a time-independent generator $L[\rho] = \dot{\rho}$ of the dynamics, this second approach leads to a simple expression for the entropy production rate:
\be\label{d:sigmaSp}
  \dot\sigma_\text{Sp} &=& -\dfrac{d}{dt}S\left(\rho_{S}(t)\Big\|\rho_\infty\right) \nonumber\\
  &=&\dfrac{d}{dt}S(t) + \text{Tr}\left( L(\rho)\ln\rho_\infty\right),
\ee
with $\rho_\infty$ a fixed point of the generator, i.e. $ L(\rho_\infty)=0$. Extension to the case of a time-dependent generator is obtained from the second line of Eq.~\eqref{d:sigmaSp}, by taking $\rho_\infty$ to be the instantaneous fixed point of the generator \cite{picatoste2026localapproachentropyproduction}. 

The entropy production rate defined in Eq.~\eqref{d:sigmaSp} vanishes for $\rho=\rho_\infty$ by construction. While this is natural if $\rho_\infty$ corresponds to an equilibrium state, such entropy production may miss part of the total entropy production when the steady state is a nonequilibrium steady state (e.g., if the system is driven, or if the environment is out of equilibrium). In that case, $\dot\sigma_\text{Sp} $ is typically interpreted as the nonadiabatic entropy production rate \cite{yukawa2001,Horowitz_14}. In contrast, $\dot\sigma_\text{tot}$ and its generalization to arbitrary environment state capture both nonadiabatic and adiabatic contributions to entropy production rate, and we therefore refer to it as the total entropy production below, while we refer to $\dot\sigma_\text{Sp}$ as the Spohn entropy production rate. \\

Note that in both approaches, positivity of the entropy production rate requires vanishing system-bath mutual information at time $t$. This condition is fulfilled for the, weak coupling, Markovian dynamics considered here. In contrast, information backflows can lead to violations of those two bounds. Inequality \eqref{d:sigmatot} can however be modified as $\sigma_\text{tot} \geq \Delta I_{SB}$ where $I_\text{SB}$ is the quantum mutual information between $S$ and $B$ to account for such violations \cite{Elouard_23}. Similarly, $\dot\sigma_\text{Sp}$ has been used as a witness of information backflow \cite{picatoste2026localapproachentropyproduction,Theret2026}.

We now apply those two definitions to the case of the GAME. In Eq.~\eqref{total_entropy}, we can directly inject the bath energy flow predicted by the GAME, Eq.~\eqref{eq:dEbGAME}, to obtain:
\begin{equation}\label{eq:sigmaSpGAME}
\frac{dS}{dt}+\beta\left(-i\Tr_{S}\left(H_{S}\left[H_{LS},\rho_{S}\right]\right)-\Tr_{S}\left(H_{S}D^{G}(\rho_{S})\right)\right).
\end{equation}

Implementing the second approach requires to derive the stationary state of the GAME. This can be done analytically by using a perturbative treatment \cite{Thingna_12,Thingna_13,Lee_22,Lobjeko_24}. Indeed as we mentioned in the previous section, the GAME is a first order approximation of the real dynamics with respect to $\gamma$ (Eq.~\eqref{me_expansion}). Therefore the stationary state will be accurate at best up to the first order and we can approximate it as $\rho_{\infty}\approx\rho_{\infty}^{(0)}+\gamma\rho_{\infty}^{(1)}$. Then we can plug the steady state in the GAME and we get the following consistent equations

\begin{widetext}
    \begin{equation}\label{exp_st}
    \begin{split}
        L^{(0)}(\rho^{(0)}_{\infty})&=-i[H_{S},\rho^{(0)}_{\infty}]=0,\\
        L^{(0)}(\rho_{\infty}^{(1)})+L^{(1)}(\rho_{\infty}^{(0)})&=-i[H_{S},\rho^{(1)}_{\infty}]-i[H_{LS},\rho_{\infty}^{(0)}]+D^{G}(\rho_{\infty}^{(0)})=0.
    \end{split}
\end{equation}

\end{widetext}

From the first equation we deduce that the 0th order term is diagonal in the eigenbasis of the system Hamiltonian and thus we can write it as $\rho_{\infty}^{(0)}=\sum_{n}p_{n}^{(0)}\ket{n}\bra{n}$. We can then exploit its simple structure to determine its expression with the second equation. For that, we simply project it in the diagonal elements of $H_{S}$ (therefore the commutator terms vanish) to get $\bra{n}D^{G}(\rho_{_\infty}^{(0)})\ket{n}=0$. In the end one can show that the 0th order term is given by the Gibbs state with respect to $H_{S}$ i.e $\rho_{\infty}^{(0)}=\exp(-\beta H_{S})/Z_{S}$ (see App.~\ref{appC} or Refs.~\cite{Thingna_12,Lobjeko_24} for more details). The key point to end up with this result is that the decay rates satisfy the Kubo-Martin-Schwinger (KMS) condition $\gamma(-\omega)=e^{-\beta\omega}\gamma(\omega)$.

With this result in mind, we can give a more detailed form of the steady state. Indeed since the 0th order term is given by the the Gibbs state with $H_{S}$, we can write the exact steady state as a Gibbs state of an effective Hamiltonian $H_\text{eff}$, that can be expanded with respect to $\gamma$ and for which the 0th order term is $H_{S}$
\begin{equation}\label{ss}
    \rho_{\infty}\propto e^{-\beta H_\text{eff}}= e^{-\beta(H_{S}+\sum_{n=1}\gamma^{n}H^{(n)})}\approx e^{-\beta(H_{S}+\gamma H^{(1)})}
\end{equation}

For the GAME, it is sufficient to go up to first order $H_\text{eff}\approx H_{S}+\gamma H^{(1)}$. Following the procedure proposed in Ref.~\cite{Lobjeko_24}, we can relate the Hamiltonian correction term $H^{(1)}$ to $\rho_{\infty}^{(1)}$. We assume that the eigenoperators can span the entire Hilbert space of the system and use them to expand $H^{(1)}$ as

\begin{equation}
    H^{(1)}=\sum_{\overline{\omega}}\sum_{\omega,\omega'\in\mathcal{L}(\overline{\omega})}H^{(1)}(\omega,\omega')A^{\dagger}(\omega')A(\omega).
\end{equation}

Then by expanding the steady state we get (see App.~\ref{appC} for details)

\begin{equation}\label{rhoinf1}
\begin{split}
    \rho_{\infty}^{(1)}=&-\sum_{\overline{\omega}}\sum_{\omega,\omega'\in\mathcal{L}(\overline{\omega})}H^{(1)}(\omega,\omega')\alpha(\omega'-\omega) \rho_{\infty}^{(0)}\left(A^{\dagger}(\omega')A(\omega)\right.\\
    &\left.-\Tr\left(\rho_{\infty}^{(0)}A^{\dagger}(\omega')A(\omega)\right)\right),
\end{split}
\end{equation}
where $\alpha(x)=(e^{\beta x}-1)/x$. Finally we can use the 1st order consistent relation (second line of Eq.~\eqref{exp_st}) to get the explicit expression of the correction Hamiltonian. We showed that it is given by (see App.~\ref{appC})

\begin{equation}
    \begin{split}
        H^{(1)}&=H_{LS}+K,\\
        K&=\sum_{\overline{\omega}}\sum_{\omega,\omega'\in\mathcal{L}(\overline{\omega})}K(\omega,\omega')A^{\dagger}(\omega')A(\omega),\\
K(\omega,\omega')&=\frac{i\sqrt{\gamma(\omega')\gamma(\omega)}\tanh(\beta(\omega-\omega')/4)}{2}.
    \end{split}
\end{equation}

Let us briefly mention that actually the consistent relations given by Eq.~\eqref{exp_st} only allow us to get the off-diagonal terms of the correction Hamiltonian (since it involves commutation relations between $\rho_{\infty}^{(1)}$ and $H_{S}$). In principle one would need to get the dynamical equation of the system at the second order in $\gamma$ to get the diagonal terms \cite{Mori_08,Flemming_11}. Otherwise it has been shown that an analytical continuation of the off-diagonal terms of $H^{(1)}$ can be done instead \cite{Thingna_12} (in our case we get $\lim_{\omega'\rightarrow\omega}K(\omega,\omega')=0$ and $\lim_{\omega'\rightarrow\omega}H^{(1)}(\omega,\omega')=S(\omega)$). 

But in the end, the knowledge of the diagonal terms does not matter in our case as they do not play any role in the thermodynamic description of the GAME. Indeed the Spohn entropy production is given by

\begin{equation}\label{eq:spohnGAME}
    \dot\sigma_\text{Sp}(t)=\frac{dS}{dt}-\beta\left(i\Tr_{S}\left([K,\rho_{S}]H_{S}\right)+\Tr_{S}\left(H_{S}D^{G}(\rho_{S})\right)\right).
\end{equation}

We got rid of the terms that are $\mathcal{O}(\gamma^{2})$, even though they are obtained with the Spohn inequality and contribute to the positivity of $\dot\sigma_\text{Sp}$. Indeed those terms are negligible in the regime of validity of the GAME and a higher order description would be necessary to treat them rigorously. \\

\subsection{Transient coherent contributions to entropy production}\label{sec.IIIC}

By comparing Eqs.~\eqref{dEB2} and \eqref{total_entropy} to Eq.~\eqref{eq:spohnGAME}, we clearly see that the two entropy production rates give different values for the GAME. Namely,
\begin{equation}
\begin{split}
    \Delta\dot\sigma(t)&=\dot\sigma_\text{tot}(t)-\dot\sigma_\text{Sp}(t)=-i\beta\Tr_{S}\left([H_{LS}-K,\rho_{S}]H_{S}\right).
\end{split}
\end{equation}

Such difference is reminiscent to the case of a system in a non-equilibrium steady state (NESS), where $\dot\sigma_\text{Sp}$ and $\Delta\dot\sigma$ play respectively the roles of effective nonadiabatic and adiabatic entropy productions \cite{yukawa2001,Horowitz_14}. A NESS typically occurs when the system is driven or coupled to multiple reservoirs. The adiabatic entropy production then corresponds to the part of the total entropy production that accounts for keeping the system in the NESS, while the nonadiabatic entropy production is generated when relaxing towards the NESS \cite{Esposito_2010b,Esposito_2010c}. 
Here, the discrepancy between both entropy production rates comes from the fact that the steady state of the GAME is not a Gibbs state with respect to $H_{S}$, but rather with respect to the operator $H_\text{eff} = H_{S}+\gamma H^{(1)}$. Notably, it contains coherence in the basis of $H_{S}$ induced by the Lamb-shift and the nonsecular jumps of the GAME generator Eq.~\eqref{GAME}.

However, in contrast with the case of a NESS, the discrepancy between both entropy production rates is purely transient. Indeed, we have at the leading order $\lim_{t\rightarrow\infty}\Delta\dot\sigma(t)=-i\beta\Tr_{S}\left([H_{LS}-K,\rho^{(0)}_{\infty}]H_{S}\right)=0$. 
In the non-driven, single-bath setting considered here, the stationary state exhibits the expected equilibrium properties: system observables are constant, while energy currents and both entropy-production rates vanish to the perturbative order retained. Thermal equilibrium of an interacting open system does not require its reduced state to be Gibbsian with respect to the bare Hamiltonian $H_{S}$; in particular, stationary coherences in its eigenbasis do not by themselves imply a non-equilibrium steady state or a sustained dissipation cost.
The transient nature of $\Delta\dot\sigma$ will be clearer when looking the case of multiple baths (Sec.~\ref{sec.IIIG}). 

We finally mention that, while the total and Spohn entropy production rates are always positive, it is not necessarily the case for $\Delta\dot\sigma$. This was already shown to be the case for quantum NESS \cite{Manzano_18} where the adiabatic entropy production does not necessarily fulfill a second law.

\subsection{Recovering the 2nd law for the local regime}\label{sec.IIID}

The total entropy production for the LNME can be directly obtained by passing to the limit of vanishing $\delta H_S$ from Eq.~\eqref{eq:sigmaSpGAME}. We get:
\be
    \dot\sigma_\text{tot}^{(0)}(t)&=&\frac{dS^{(0)}}{dt}+\beta\frac{dE^{(0)}_{B}}{dt}\\
    &=&\frac{dS^{(0)}}{dt}-\beta\Tr_{S}\left(H^{(0)}_{S}D^{(0)}(\rho_{S})\right)\\
    &\equiv&\frac{dS^{(0)}}{dt}-\beta \dot{Q}_{S}^{(0)},
\label{dStot_local}
\ee
where $S^{(0)}$ is the von Neumann entropy obtained with the LNME. The entropy production therefore takes a form similar to that of the secular Linbdlad equation. At this order, the change of entropy of the bath is given by the total change of energy of the system that we can completely identify with $Q^{(0)}$.

Finally, let us see how $\dot\sigma_\text{Sp}$ and $\Delta\dot\sigma$ change for the LNME. As before, we take the limit where all the Bohr frequencies of the same cluster tend to $\overline\omega$. Noting that $\lim_{(\omega,\omega')\rightarrow(\overline\omega,\overline\omega)}K(\omega,\omega')=0$ and $H^{(1,0)}=H_{LS}^{(0)}$, the Spohn entropy production simply becomes

\begin{equation}
    \dot\sigma_\text{Sp}^{(0)}(t)=\frac{dS^{(0)}}{dt}-\beta\Tr_{S}\left(H^{(0)}_{S}D^{(0)}(\rho_{S})\right)=\dot\sigma^{(0)}_\text{tot}(t).
\end{equation}
Therefore, both the Spohn and total entropy productions become equal (and no effective adiabatic contribution remains) suggesting that the system relaxes to an equilibrium state where the potential presence of coherence in the basis of $H_{S}^{(0)}$ does not matter, i.e. leads to a negligible contribution to heat flow and entropy production. This equilibrium state can be identified as for the GAME from a perturbative treatment but now treating both $\gamma$ and the magnitude $\Omega$ of the frequency differences inside a given cluster (or equivalently, the magnitude of $\delta H_{S}$) as expansion parameters. 
The 0th order with respect to $\gamma$, then splits as $L^{(0)}(\rho)=L^{(0,0)}(\rho)+\Omega L^{(0,1)}(\rho)$ with $L^{(0,0)}(\rho)=-i\left[H_{S}^{(0)},\rho\right]$ and $ L^{(0,1)}(\rho)=-i\left[\delta H_{S},\rho\right]$ and the first order term as $L^{(1)}(\rho)=\sum_{n}\Omega^{n}L^{(1,n)}(\rho)$. In the LNME, we keep only terms $L^{(0,1)}$ and $L^{(1,0)}$; it is therefore valid when $\Omega$ is at most of the same order of $\gamma$. Expanding the steady state as $\rho_{\infty}\approx\rho_{\infty}^{(0,0)}+\Omega\rho_{\infty}^{(0,1)}+\gamma\rho_{\infty}^{(1,0)}$, the consistent relations obtained for the steady state of the GAME (Eq.~\eqref{exp_st}) become for the LNME

\begin{widetext}
    \begin{equation}\label{eq:steady-lnme}
    \begin{split}
        L^{(0,0)}(\rho^{(0,0)}_{\infty})&=-i[H_{S}^{(0)},\rho^{(0,0)}_{\infty}]=0,\\
        L^{(0,0)}(\rho_{\infty}^{(0,1)})+L^{(0,1)}(\rho_{\infty}^{(0,0)})&=-i\left[H_{S}^{(0)},\rho_{\infty}^{(0,1)}\right]-i\left[\delta H_{S},\rho_{\infty}^{(0,0)}\right]=0\\
        L^{(0,0)}(\rho_{\infty}^{(1,0)})+L^{(1,0)}(\rho_{\infty}^{(0,0)})&=-i[H^{(0)}_{S},\rho^{(1,0)}_{\infty}]+D^{(0)}(\rho_{\infty}^{(0,0)})=0.
    \end{split}
\end{equation}

\end{widetext}

From the first relation we deduce that $\rho_{\infty}^{(0,0)}$ is diagonal in the eigenbasis of $H_{S}^{(0)}$, and like for the GAME, we can project the third relation in this basis to get an explicit expression. 
As we have chosen $[H_S^{(0)},\delta H_S]=0$, the eigenoperators $A(\omega)$ can be chosen to be simultaneous eigenoperators of $H_S^{(0)}$ and $\delta H_S$. This implies that the LNME jump operators $A_{\bar\omega}^{(0)}$ are eigenoperators of $H_{S}^{(0)}$. Assuming this is the case and using the KMS condition $\left(\gamma(-\overline\omega)=e^{-\beta\overline{\omega}}\gamma(\overline\omega)\right)$ we can deduce that $\rho_{\infty}^{(0,0)}=e^{-\beta H_{S}^{(0)}}/Z_{S}^{(0)}$. Like for the secular Lindblad equation, the state $\rho_{\infty}^{(1,0)}$ is not fully determined by the LNME generator only, and one would need to include terms at second order in $\gamma$ to obtain the correction to the populations at this order. Similarly, in $\rho_{\infty}^{(0,1)}$, the off-diagonal terms can be obtained from the second equality in Eq.~\eqref{eq:steady-lnme},  but not the diagonal terms, which would be determined by Liouvillian terms of order $\mathcal{O}(\gamma\Omega)$. Nevertheless, the exact knowledge of these terms in not necessary to derive the expression for the Spohn entropy production from its inequality (just as for the secular Lindblad equation). It is straightforward to see that we have

\begin{equation}
    \dot\sigma_\text{Sp}^{(0)}(t)=-\frac{dS(\rho_{S}(t)||\rho_{\infty}^{(0,0)})}{dt}
\end{equation}

This result is in full agreement with Ref.~\cite{Potts_2021}, where the authors derive the entropy production from the Spohn inequality with $\rho_{\infty}$ taken equal to the Gibbs state constructed with a thermodynamic bookkeeping Hamiltonian (which in fact corresponds to $H_S^{(0)}$). Additionally, here we outline that this Gibbs state is the correct steady state at the order of magnitude captured by the LNME, and both $\dot\sigma$ and $\dot\sigma_\text{Sp}$ coincide in this regime.

\subsection{First law and work extraction}\label{sec.IIIE}

Here we briefly discuss the implication of our thermodynamic description of the GAME on the first law. The main difference with the usual case of fully secular master equations is the nontrivial role of the coupling energy and the Lamb-shift Hamiltonian. Following the approach of Ref.~\cite{Esposito_2010}, the heat can be identified with the opposite energy change of the bath, henceforth rewriting the total entropy production in the Clausius form:
\begin{eqnarray}
    \dot\sigma_\text{tot} = \dfrac{d}{dt}S -\beta\dot Q,\quad\dot Q \equiv -\dfrac{dE_B}{dt}.
    \label{heat_esposito}
\end{eqnarray}
From the energy balance \eqref{eq:en_bal}, the first law then reads:
\begin{eqnarray}
   \dfrac{dE_S}{dt} &=& \dot Q + \dot W, \quad \dot W \equiv -\dfrac{dE_I}{dt}. 
\end{eqnarray}
The identification of the work with the variation of the interaction energy is physically justified given that it corresponds to the energy input from the operator switching on and off the coupling to the bath at the beginning and the end of an isothermal transformation. In this sense, when including such isothermal transformation in a thermodynamic cycle, this amount of energy constitutes an additional resource coming neither from the initial state of the bath or the system. Another view, more adapted to the case of a always-on system-bath interaction, is that the variation of the coupling energy over time towards is a resource which can be extracted from (or which has been paid to prepare) the nonequilibrium initial condition. In either prescriptions, the second law Eq.~\eqref{total_entropy} prevent any cyclical work extraction from the single bath via 
\begin{equation}
  W=-Q \geq 0\quad\text{(over a cycle),} 
\end{equation}
and because $\underset{t\to\infty}{\lim} \dot W =0$.

It is also interesting to mention that since the system relaxes to a steady state that contains coherence with respect to the system  Hamiltonian, one could extract work out of it. However, this amount is bounded and thus a (at least) higher energetic price is needed to pay in order to extract it. The maximum amount of extractable work from a quantum state can be evaluated with the so-called ergotropy defined as $R(\rho)=\Tr_{S}(H_{S}\rho)-\min_{U}\Tr_{S}(H_{S}U\rho_{P}U^{\dagger})=\Tr_{S}(H_{S}\rho)-\Tr_{S}(H_{S}\rho_{P})$ where $\rho_{P}$ is the passive state of the system \cite{Allahverdyan_2004}. By integrating the second law (Eq.~\eqref{d:sigmatot}) one can show that the ergotropy for the steady state is bounded by (see App.~\ref{appD})

\begin{equation}
    R(\rho_{\infty})\leq-\Delta E_{I}+F_{\beta}(\rho_{S}(0))-F_{\beta}(\rho_{\beta}),
\end{equation}
where $F_{\beta}(\rho)=\Tr_{S}(H_{S}\rho)-\beta^{-1}S(\rho)$ is the non-equilibrium free energy, and we denoted $\rho_{\beta}$ the Gibbs state at inverse temperature $\beta$ with respect to the system Hamiltonian. The bound thus accounts for two contributions: the first one coming from the energy cost for switching on the coupling. The second one given by the change of free energy, accounts for preparing initially the system in a non-thermal state that can contain resources. Since we do not realize a cycle but just a single transformation, we do not pay an additional cost for returning the system from the passive state to the initial state and this explains why a larger amount of energy than the coupling contribution could be extracted.

\subsection{Case of multiple baths}\label{sec.IIIG}

Here we extend our results to the case where the system is coupled to $N_{B}>1$ thermal baths that are at different temperatures. The coupling Hamiltonian is assumed to be given by $H_{I}=\sum_{\alpha=1}^{N_{B}}A_{\alpha}\otimes B_{\alpha}=\sum_{\alpha=1}^{N_{B}}H_{I}^{\alpha}$, $\alpha$ labeling the baths, and the density matrix of each of them is given by $\rho_{B_{\alpha}}=e^{-\beta_{\alpha}H_{B_{\alpha}}}/Z_{B_{\alpha}}$. Assuming that the bath are uncorrelated, we find that most of the results derived for a single bath can be straightforwardly extended to this case. The GAME is then simply obtained by adding the different contributions from each bath

\begin{equation}\label{GAME2}
    \begin{split}
        \frac{d\rho_{S}}{dt}&=-i\left[H_{S}+\sum_{\alpha=1}^{N_{B}}H^{\alpha}_{LS},\rho_{S}\right]+\sum_{\alpha=1}^{N_{B}}D_{\alpha}^{G}(\rho_{S}),\\
        H^{\alpha}_{LS}&=\sum_{\overline{\omega}}\sum_{\omega,\omega'\in\mathcal{L}(\overline{\omega})}S_{\alpha}(\omega,\omega')A_{\alpha}^{\dagger}(\omega')A_{\alpha}(\omega),\\
        D_{\alpha}^{G}(\rho_{S})&=\sum_{\overline{\omega}}\sum_{\omega,\omega'\in\mathcal{L}(\overline{\omega})}\sqrt{\gamma_{\alpha}(\omega)\gamma_{\alpha}(\omega')}\left(A_{\alpha}(\omega)\rho_{S} A_{\alpha}^{\dagger}(\omega')\right.\\
        &-\left.\frac{1}{2}\left\{A_{\alpha}^{\dagger}(\omega')A_{\alpha}(\omega),\rho_{S}\right\}\right).
    \end{split}
\end{equation}

The energy balance takes the form
\begin{equation}
    \begin{split}
        &\frac{dE_{S}}{dt}+\sum_{\alpha=1}^{N_{B}}\frac{dE_{B_{\alpha}}}{dt}+\frac{dE^{\alpha}_{I}}{dt}=0,\\
&\frac{dE_{S}}{dt}=\sum_{\alpha=1}^{N_{B}}-i\Tr_{S}\left(H_{S}\left[H^{\alpha}_{LS},\rho_{S}\right]\right)+\Tr_{S}\left(H_{S}D_{\alpha}^{G}(\rho_{S})\right),\\
&\frac{dE_{B_{\alpha}}}{dt}=-i\Tr_{S}\left(H_{S}\left[H^{\alpha}_{LS},\rho_{S}\right]\right)-\Tr_{S}\left(H_{S}D_{\alpha}^{G}(\rho_{S})\right),\\
&\frac{dE^{\alpha}_{I}}{dt}=2i\Tr_{S}\left(H_{S}\left[H^{\alpha}_{LS},\rho_{S}\right]\right).
    \end{split}
\end{equation}

The total entropy production is given by the variation of the vN entropy of the system and each bath

\begin{equation}
        \dot\sigma_\text{tot}(t)=\frac{dS}{dt}-\sum_{\alpha=1}^{N_{B}}\beta_{\alpha}\frac{dQ_{B_{\alpha}}}{dt}\geq0
    \label{total_ep_multibath}
\end{equation}
where we can identify the heat provided by each bath $Q_{B_{\alpha}}$ (Eq.~\eqref{heat_esposito}). Evaluating the Spohn entropy production is less straightforward as the steady state is now a NESS, whose analytical expression can be very complex. However, the perturbative treatment can be extended, leading to insights into its structure \cite{Thingna_13}. Again we expand the steady state as $\rho_{\infty}\approx\rho^{(0)}_{\infty}+\gamma\rho^{(1)}_{\infty}$ and we can deduce that the 0 order term is diagonal in $H_{S}$, while it is not a Gibbs state in general. To know exactly the population of $\rho^{(0)}_{\infty}$, we need to solve the algebraic equation $\sum_{\alpha}\bra{n}D_{\alpha}(\rho_{\infty})\ket{n}=0$, which can be done numerically and sometimes analytically when the eigenoperators satisfy convenient properties (e.g simple anti/commutation relations).

Interestingly, we can show that two occupation populations $p^{(0)}_{n}$ and $p^{(0)}_{m}$ at two eigenstates that are separated by a frequency transition $\epsilon_{n}-\epsilon_{m}=\omega$ satisfy an effective detailed balance condition (see App.~\ref{appE} for details)

\begin{equation}
    \frac{p^{(0)}_{n}}{p^{(0)}_{m}}=e^{-\tilde \beta(\omega)\omega},
\end{equation}
where $\tilde\beta(\omega)$ is an apparent (virtual) inverse temperature that depends on the system and the bath through the system operators $A_{\alpha}$ and the decay rates $\gamma_{\alpha}(\omega)$ \cite{Latune_2019}. Therefore the 0 order term describes a state where each transition in the system is locally coupled and in thermal equilibrium with a bath at the inverse temperature $\tilde\beta(\omega)$.  
Additionally from the KMS condition, we can show that $\tilde{\beta}(\omega)=\tilde{\beta}(-\omega)\geq0$ for all frequency transitions (see App.~\ref{appE}).

We can also determine the 1st order correction to the steady state if the system by assuming that it can be expanded in terms of the eigenoperators of $H_S$ as 

\begin{equation}
    \begin{split} \rho_{\infty}^{(1)}=&\sum_{\alpha}\sum_{\overline{\omega}}\sum_{\omega,\omega'\in\mathcal{L(\overline{\omega})}}R_{\alpha}(\omega,\omega')\rho_{\infty}^{(0)}\left(A^{\dagger}_{\alpha}(\omega')A_{\alpha}(\omega)\right.\\
        &\left.-\Tr_{S}\left(\rho_{\infty}^{(0)}A^{\dagger}_{\alpha}(\omega')A_{\alpha}(\omega)\right)\right).
    \end{split}
\end{equation}

The expression of the coefficient $R_{\alpha}(\omega,\omega')$ is provided in App.~\ref{appE}. We see that the 1st order correction to the steady state can be written as a sum of independent contributions due to each bath. This is a consequence of the dissipator and the Lamb-shift terms being additive, which, combined with the Spohn inequality formulated for each generator $D^G_\alpha$, leads to

\begin{equation}
\begin{split}
    \dot\sigma_\text{Sp}(t)=&\frac{dS}{dt}-i\Tr_{S}\left(\left[H_{S},\rho_{S}\right]s^{(1)}\right)\\
    &-i\sum_{\alpha}\Tr_{S}\left(\left[H^{\alpha}_{LS},\rho_{S}\right]\log\left(\rho^{(0)}_{\infty}\right)\right)\\
    &+\sum_{\alpha}\Tr_{S}\left(D^{G}_{\alpha}(\rho_{S})\log\left(\rho^{(0)}_{\infty}\right)\right)\geq 0, 
\end{split}
\end{equation}
where $s^{(1)}$ is the first order Taylor expansion of $\log\left(\rho_{\infty}^{(0)}+\gamma\rho_{\infty}^{(1)}\right)$ (see App.~\ref{appE}). Again we observe transient contributions due to the Lamb-shift and the 1st order correction of the steady state (second and third term). The last term is analogous to the excess heat in classical thermodynamics i.e the part of the heat exchanged with the baths 
when the system is relaxing towards a NESS \cite{Oono_98,Saito_11,Yuge_13}. Finally we can deduce the difference between both entropy production rates

\begin{equation}
\begin{split}
    \Delta\dot\sigma(t)=&i\Tr_{S}\left(\left[H_{S},\rho_{S}\right]s^{(1)}\right)\\
    &-\sum_{\alpha}i\beta_{\alpha}\Tr_{S}\left(\left[H^{\alpha}_{LS},\rho_{S}\right]\left(H_{S}-\beta^{-1}_{\alpha}\log\left(\rho^{(0)}_{\infty}\right)\right)\right)\\&-\sum_{\alpha}\beta_{\alpha}\Tr_{S}\left(D^{G}_{\alpha}\left(\rho_{S}\right)\left(H_{S}+\beta^{-1}_{\alpha}\log\left(\rho^{(0)}_{\infty}\right)\right)\right).
\end{split}
\end{equation}

As for the case of one thermal bath, the first two terms are two transient contributions related to the  Lamb-shift and the nonsecular transitions which deform the steady state with respect to the Gibbs steady state of the secular Lindblad equation. The third term corresponds to the housekeeping heat \cite{Oono_98,Manzano_18}, i.e. the heat flow remaining when the system is in a NESS. Its absence in the single bath case was consistent with the steady state being a genuine equilibrium state in that case. In contrast, the temperature bias between several baths typically cause the steady state to be a NESS, associated with continuous entropy production and heat flows through the system. Namely, at steady state, the two first contributions vanish, so as the Spohn entropy production, and we have at leading order

\begin{eqnarray}
\lim_{t\rightarrow\infty}\dot\sigma_\text{tot}(t)&=&\lim_{t\rightarrow\infty}\Delta\dot\sigma(t)\nonumber\\
&=&-\sum_{\alpha}\beta_{\alpha}\Tr_{S}\left(D^{G}_{\alpha}\left(\rho^{(0)}_{\infty}\right)H_{S}\right)\geq 0,\quad
\label{entropyNESS}
\end{eqnarray}
which corresponds to the so-called adiabatic entropy production. Let us notice that we can identify in the sum, the stationary heat flux from each thermal bath.


\section{Examples}\label{sec.IV}

\begin{figure*}[!t]
  \centering
  \includegraphics[width=0.9\textwidth]{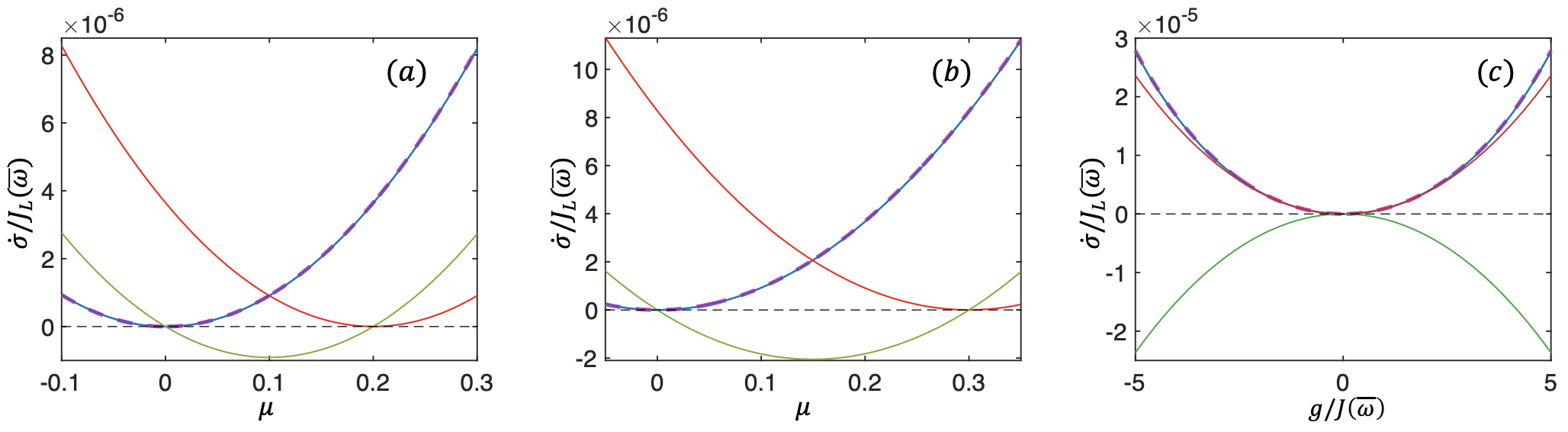}
  \caption{Entropy production rate calculated for the parametric state (Eq.~\eqref{eq:param}). The blue and dashed purple curves correspond respectively to the total and Spohn entropy productions obtained with GAME. The green curve corresponds to the naive entropy production and the red one to the correct entropy production for the LNME. The parameters are $\overline{\omega}=50J(\overline{\omega})$, $\delta=10^{-3}\overline{\omega}$, $\beta\overline{\omega}=10$, $c=0.02$ and $\Lambda=10\overline{\omega}$. (a) Entropy production as function of $\mu$ with $g=0.02\overline{\omega}$. (b) Entropy production as function of $\mu$ with $g=0.03\overline{\omega}$. (c) Entropy production as function of $g$ with $\mu=\mu^{*}=\beta g/2$.}
  \label{Fig5A1}
\end{figure*}

\subsection{Two interacting qubits coupled to a thermal bath}

In this first example, we consider two interacting qubits, labeled $L$ (left) and $R$ (right), and we suppose that the left oscillator is coupled to a bosonic thermal bath at the inverse temperature $\beta$. The total Hamiltonian is given by

\begin{equation}
\begin{split}
    H_{S}&=\omega_{L}\sigma^{+}_{L}\sigma^{-}_{L}+\omega_{R}\sigma^{+}_{R}\sigma^{-}_{R}+g\left(\sigma^{+}_{L}\sigma^{-}_{R}+\sigma^{-}_{L}\sigma^{+}_{R}\right),\\
    H_{B}&=\sum_{n}\omega_{n}b^{\dagger}_{n}b_{n},\\
    H_{I}&=A_{L}\sum_{n}c_{n}\left(b^{\dagger}_{n}+b_{n}\right),\; A_{L}=\sigma^{+}_{L}+\sigma^{-}_{L}
\end{split}
\end{equation}

where $\sigma^{+}_{\alpha}=(\sigma^{-}_{\alpha})^{\dagger}=\ket{1_{\alpha}}\bra{0_\alpha}$ and $\ket{1_{\alpha}}$ ($\ket{0_{\alpha}}$) is the excited (ground) state of a qubit. We assume that the inter-qubit coupling is weak i. e $g\ll\omega_{\alpha}$. By diagonalizing the system Hamiltonian, we can identify its spectral structure

\begin{equation}
\begin{split}
H_{S}&=\omega_{+}\ket{+}\bra{+}+\omega_{-}\ket{-}\bra{-}+2\overline{\omega}\ket{1_{L}1_{R}}\bra{1_{L}1_{R}},\\
\ket{+}&=\sin(\theta)\ket{0_{L}1_{R}}+\cos(\theta)\ket{1_{L}0_{R}}\\
    \ket{-}&=\cos(\theta)\ket{0_{L}1_{R}}-\sin(\theta)\ket{1_{L}0_{R}},\\
    \omega_{\pm}&=\overline{\omega}\pm\sqrt{\frac{\delta^{2}}{4}+g^{2}}=\overline{\omega}\pm\Omega,\\
    \overline{\omega}&=\frac{\omega_{R}+\omega_{L}}{2},\hspace{0.2 cm}\delta=\omega_{L}-\omega_{R},\hspace{0.2 cm}\tan(2\theta)=\frac{2g}{\delta}.    
\end{split}
\end{equation}

We focus on the regime where the detuning is small enough such that $\Omega\ll\overline\omega$. Therefore we can identify $H_{S}^{(0)}$ and $\delta H_{S}$ as 
\begin{equation}
    \begin{split}
        H^{(0)}_{S}&=\overline{\omega}\left(\ket{+}\bra{+}+\ket{-}\bra{-}+2\ket{1_{L}1_{R}}\bra{1_{L}1_{R}}\right)\\
        &=\overline{\omega}\left(\sigma^{+}_{L}\sigma^{-}_{L}+\sigma^{+}_{R}\sigma^{-}_{R}\right),\\
        \delta H_{S}&=\Omega\left(\ket{+}\bra{+}-\ket{-}\bra{-}\right).
    \end{split}
\end{equation}

The Bohr frequencies of the dominant Hamiltonian are $+\overline{\omega}$ and $-\overline{\omega}$ and we can thus deduce the corresponding clusters $\mathcal{L}(\pm\overline{\omega})=\{\pm\omega_{+},\pm\omega_{-}\}$. With the knowledge of the eigenstates of the system, we can deduce the eigenoperators that are involved in the dynamics

\begin{equation}
\begin{split}
    A_{L}&=A(\omega_{+})+A(\omega_{-})+A(-\omega_{+})+A(-\omega_{-}),\\
    A(\omega_{+})&=A^{\dagger}(-\omega_{+})=\cos(\theta	) \left( \ket{0_{L}0_{R}}\bra{+} + \ket{-}\bra{1_{L}1_{R}}\right),\\
    A(\omega_{-})&=A^{\dagger}(-\omega_{-})=\sin(\theta) \left( \ket{+}\bra{1_{L}1_{R}} - \ket{0_{L}0_{R}}\bra{-} \right).
\end{split}
\end{equation}

The GAME is derived by following the coarse-grained procedure of the system evolution by taking $\omega_{L}^{-1},\omega_{R}^{-1}\ll\Delta t\ll\Omega^{-1},\gamma^{-1}$ and using the geometric approximation. For this setting we get

\begin{equation}
\begin{split}
    &\frac{d\rho_{S}}{dt}=-i[H_{S}+H_{LS},\rho_{S}]\\
&+\sum_{\omega=\overline{\omega},-\overline{\omega}}\left(A_{\omega}\rho_{S} A_{\omega}^{\dagger}-\frac{1}{2}\{A_{\omega}^{\dagger}A_{\omega},\rho_{S}\}\right),
\end{split}
\label{GAME_qubit}
\end{equation}

where the jump operators are

\begin{equation}
\begin{split}
A_{\pm\overline{\omega}}&=\sqrt{\gamma(\pm\omega_{+})}A(\pm\omega_{+})+\sqrt{\gamma(\pm\omega_{-})}A(\pm\omega_{-})\\
    \gamma(-\omega)&= 2J(\omega)n(\omega),\hspace{0.2 cm}\gamma(\omega)= 2J(\omega)(1+n(\omega)),\\
    J(\omega)&=\pi\sum_{n}c_{n}^{2}\delta(\omega-\omega_{n}),\hspace{0.2 cm}n(\omega)=(e^{\beta\omega}-1)^{-1},
\end{split}
\end{equation}
where we introduced the spectral density function of the bath $J(\omega)$ and the Planck distribution $n(\omega)$. We choose for the bath spectral density function an Ohmic profile with a Lorentz cutoff

\begin{equation}
    J(\omega)=\frac{c\omega}{1+\left(\frac{\omega}{\Lambda}\right)^{2}},
    \label{bathSF}
\end{equation}

where $c$ is the dissipation strength and $\Lambda$ the bath cut-off.

For this example, we analyze the entropy production resulting from the dynamics with a single bath, obtained with the LNME and GAME. The LNME is simply obtained by taking the limit $\delta\Omega\rightarrow0$ for the Lamb-shift and the dissipator (see Sec.~\ref{sec.IIC} and Eq.~\eqref{LNME}). The resulting eigenoperators are $A_{\overline{\omega}}^{(0)}=\sigma^{-}_{L}$ and $A_{-\overline{\omega}}^{(0)}=\sigma^{+}_{L}$, and they are indeed local eigenoperators of $H^{(0)}_{S}$.

On top of that, we benchmark our results by calculating a "naive" entropy production. This entropy production is obtained by keeping the contribution from $\delta H_{S}$ in the heat for the LNME. It is naive in the sense that it can be derived from a non consistent derivation of the total entropy production (Eq.~\eqref{d:sigmatot}), or by writing the Spohn inequality (Eq.~\eqref{d:sigmaSp}) with $\rho_{\infty}=e^{-\beta H_{S}}/Z_{S}$ (which as we previously saw, does not correspond to the steady state). Therefore it can lead to an apparent violation of the second law \cite{Levy_2014}.

Instead of evaluating the entropy production by solving the dynamics, we perform a parametric study. We consider a generic state for the system, that is characterized by some parameters and we calculate the different entropy productions as functions of these parameters. Proceeding this way allows to reveal more precisely how the violations appear with the naive entropy production and how they are corrected for the LNME and GAME. Moreover the parametric state can actually represent a state at which the system is during the dynamics. In order to keep the analysis simple we consider a parametric state given by

\begin{equation}
    \rho_{S}(\mu)=\frac{\exp\left(-\beta H_{S}+\mu\left(\sigma^{+}_{L}\sigma^{-}_{R}+\sigma^{-}_{L}\sigma^{+}_{R}\right)\right)}{Z_{S}(\mu)},
    \label{eq:param}
\end{equation}

where $\mu$ is chosen to be real. Basically we consider states where the diagonal occupations are thermal and that contain also coherences, which is intuitively the type of state that the system would be in when relaxing. The presence of those coherences will contribute in the energy and entropy rates of the system due to the inter-qubit coupling (and so $\delta H_{S}$).

The results of our analysis are illustrated in Fig.~\ref{Fig5A1} where in panel (a) we show the different entropy production rates as function of the parameter $\mu$, with the inter-qubit coupling strength fixed at $g=0.02\overline{\omega}$. We see that the Spohn entropy (dashed purple curve) and the total entropy production (blue curve) predicted with GAME are consistently always positive, and they are very close to each other (with $\dot\sigma_{\text{Sp}}$ being slightly greater than $\dot\sigma_{\text{tot}}$). The minimum is reached for $\mu=0$ i.e when the state is given by the Gibbs state with respect to the system Hamiltonian. The minimum value for both entropies is not exactly zero but is extremely small as the contributions are $\mathcal{O}(\gamma^2)$ in that case. The entropy production for the LNME (red curve) is also consistently positive but significantly different from the latter. The minimum is reached for a different value of $\mu$. At $\mu=0$ a higher value of entropy production is predicted than GAME as $\rho_{S}(0)$ is significantly different from the steady state of the LNME $\rho^{(0,0)}_{\infty}\propto\exp\big(-\beta H^{(0)}_{S}\big)$ and introduces coherence that contribute.   

The naive entropy production (green curve) can become negative, showing apparent violation of the second law. The negative values are restricted in a given interval and the entropy becomes positive outside of it. However the interval size increases by taking higher values of $g$. This is illustrated in Fig.~\ref{Fig5A1}(b) where we have $g=0.03\overline{\omega}$ and we can see that the apparent violation of the second law occurs now for more states. Moreover the minimum reached by the naive entropy production also decreases. This is expected as the naive entropy production misses contributions that are at least $\mathcal{O}(\Omega)$. For this simple model, we can actually find analytically what is the value of $\mu$ for which the naive entropy is minimum (see App.~\ref{appF1} for the details). The minimum is reached for $\mu=\mu^{*}=\beta g/2$ and simply scales as $\dot\sigma(\mu^{*})\sim-(\beta g)^{2}$ (the exact expression is provided in App.~\ref{appF1}). The scaling is in agreement with the apparent violations reported originally for the local full secular master equation \cite{Levy_2014,Trushechkin_2016}. The profile of the entropy production evaluated at $\mu^{*}$ is shown in Fig.~\ref{Fig5A1}(c) where indeed the naive entropy production corresponds to an inverted parabola. Therefore for any finite value of $g$, the naive entropy production admits at least one state for which there is an apparent violation of the second law.   

\subsection{Two interacting harmonic oscillators each coupled to a thermal bath}

We now consider two interacting oscillators, also labeled $L$  and $R$, that are respectively coupled to a thermal bath

\begin{equation}
    \begin{split}
    H_{S}&=\omega_{R}a_{R}^{\dagger}a_{R}+\omega_{L}a_{L}^{\dagger}a_{L}+g(a_{L}a_{R}^{\dagger}+a_{L}^{\dagger}a_{R}),\\
    H_{B}&=\sum_{\alpha=L,R}\sum_{n}\omega_{\alpha,n}b_{\alpha,n}^{\dagger}b_{\alpha,n},\\
    H_{I}&=\sum_{\alpha=L,R}\sum_{n}c_{\alpha,n}\left(a_{\alpha}b_{\alpha,n}^{\dagger}+a_{\alpha}^{\dagger}b_{\alpha,n}\right),
    \end{split}
\end{equation}
where again $g\ll\omega_{L},\omega_{R}$. Once diagonalized, the Hamiltonian is given by

\begin{equation}
\begin{split}
H_{S}&=\omega_{+}a_{+}^{\dagger}a_{+}+\omega_{-}a_{-}^{\dagger}a_{-},\\
a_{+}&=\cos(\theta)a_{L}+\sin(\theta)a_{R},\\
    a_{-}&=\cos(\theta)a_{R}-\sin(\theta)a_{L},\\
    \omega_{\pm}&=\overline{\omega}\pm\sqrt{\frac{\delta^{2}}{4}+g^{2}}=\overline{\omega}\pm\Omega,\\
    \overline{\omega}&=\frac{\omega_{R}+\omega_{L}}{2},\hspace{0.2 cm}\delta=\omega_{L}-\omega_{R},\hspace{0.2 cm}\tan(2\theta)=\frac{2g}{\delta}.    
\end{split}
\end{equation}

Similarly to the previous example, we can identify $H_{S}^{(0)}=\overline\omega\left(a_{+}^{\dagger}a_{+}+a_{-}^{\dagger}a_{-}\right)=\overline{\omega}\left(a_{L}^{\dagger}a_{L}+a_{R}^{\dagger}a_{R}\right)$ with the Bohr frequencies $+\overline\omega$ and $-\overline\omega$, and the corresponding clusters are $\mathcal{L}(\pm\overline\omega)=\{\pm\omega_{+},\pm\omega_{-}\}$. We have $\delta H_{S}=\Omega\left(a_{+}^{\dagger}a_{+}-a_{-}^{\dagger}a_{-}\right)$.  We then identify the eigenoperators that are involved in the dynamics. For the left oscillator we have

\begin{equation}
    \begin{split}
        &a_{L}=A_{L}(\omega_{+})+A_{L}(\omega_{-}),\\
        &A_{L}(\omega_{+})=A_{L}^{\dagger}(-\omega_{+})=\cos(\theta)a_{+},\\
        &A_{L}(\omega_{-})=A_{L}^{\dagger}(-\omega_{-})=-\sin(\theta)a_{-},\\
    \end{split}
\end{equation}
and for the right oscillator
\begin{equation}
    \begin{split}
        &a_{R}=A_{R}(\omega_{+})+A_{R}(\omega_{-}),\\
        &A_{R}(\omega_{+})=A_{R}^{\dagger}(-\omega_{+})=\sin(\theta)a_{+},\\
        &A_{R}(\omega_{-})=A_{R}^{\dagger}(-\omega_{-})=\cos(\theta)a_{-}.\\
    \end{split}
\end{equation}

The GAME for this model is given by

\begin{equation}
\begin{split}
    &\frac{d\rho_{S}}{dt}=-i[H_{S}+H_{LS},\rho_{S}]\\
&+\sum_{\alpha=L,R}\sum_{\omega=\overline{\omega},-\overline{\omega}}\left(A_{\alpha,\omega}\rho_{S} A_{\alpha,\omega}^{\dagger}-\frac{1}{2}\{A_{\alpha,\omega}^{\dagger}A_{\alpha,\omega},\rho_{S}\}\right),
\end{split}
\label{GAME_oscillators}
\end{equation}

where the jump operators are

\begin{equation}
\begin{split}
A_{\alpha,\pm\overline{\omega}}&=\sqrt{\gamma_{\alpha}(\pm\omega_{+})}A_{\alpha}(\pm\omega_{+})+\sqrt{\gamma_{\alpha}(\pm\omega_{-})}A_{\alpha}(\pm\omega_{-})\\
    \gamma_{\alpha}(-\omega)&= 2J_{\alpha}(\omega)n_{\alpha}(\omega),\hspace{0.2 cm}\gamma_{\alpha}(\omega)= 2J_{\alpha}(\omega)(1+n_{\alpha}(\omega)),\\
    J_{\alpha}(\omega)&=\pi\sum_{n}c_{\alpha,n}^{2}\delta(\omega-\omega_{\alpha,n}),\hspace{0.2 cm}n_{\alpha}(\omega)=(e^{\beta_{\alpha}\omega}-1)^{-1}.
\end{split}
\end{equation}

\begin{figure*}[!t]
  \centering
  \includegraphics[width=0.9\textwidth]{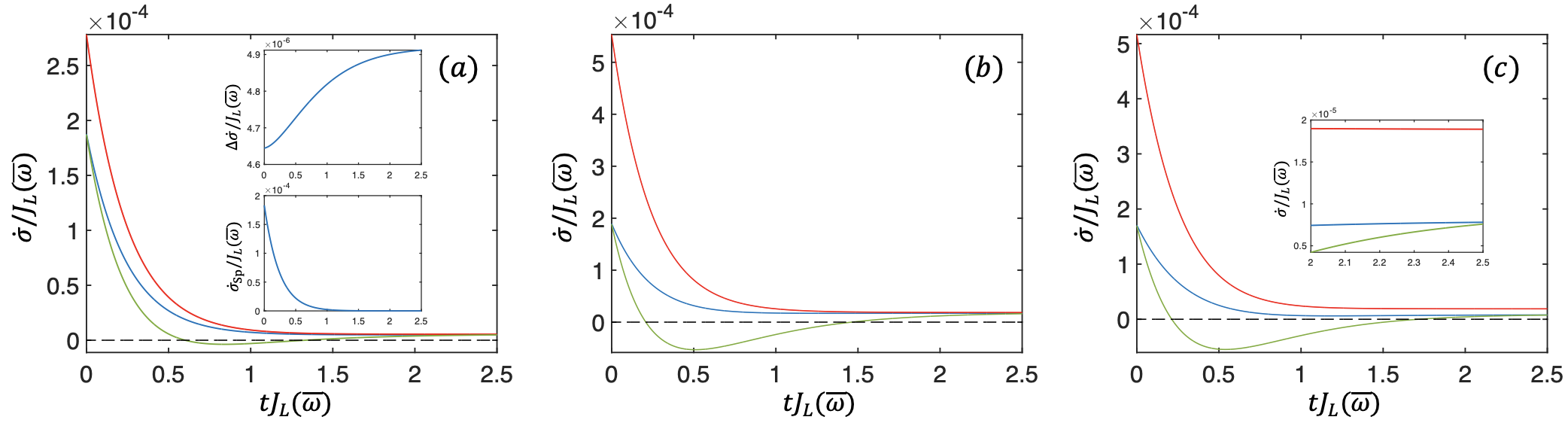}
  \caption{Total entropy production as a function of time. The blue curves correspond to the entropy obtained with GAME, the red ones to the correct entropy for the LNME and the green ones to the naive entropy production. The parameters are $\overline{\omega}=50J_{L}(\overline\omega)$, $\delta=10^{-3}\overline{\omega}$, $\beta_{R}\overline{\omega}=1$, $\beta_{L}=\beta_{R}/0.99$, $c_{L}\Lambda=0.5\overline{\omega}$ such that $\Sigma_{L}(\overline{\omega})=2.5J_{L}(\overline{\omega})$, and $\mu=10^{-2}\beta_{L}g$. The panel (a) shows the results obtained with $c_{R}=c_{L}$ and $g=5*10^{-3}\overline{\omega}$. The top and bottom insets show respectively $\Delta\dot\sigma$ and $\dot\sigma_{\text{Sp}}$. Panel (b) corresponds to $c_{R}=c_{L}$ and $g=10^{-2}\overline{\omega}$. Panel (c) corresponds to $c_{R}=0.9c_{L}$ and $g=10^{-2}\overline{\omega}$. The inset zooms on the asymptotic value of the three entropy production rates.}
  \label{Fig5B1}
\end{figure*}

The full expression of the Lamb-shift is

\begin{equation}\label{LSOscillator}
    \begin{split}
        H_{LS}&=\left(\sin^{2}(\theta)\Sigma_{L}(\omega_{-})+\cos^{2}(\theta)\Sigma_{R}(\omega_{-})\right)a_{-}^{\dagger}a_{-}\\
    &+\left(\cos^{2}(\theta)\Sigma_{L}(\omega_{+})+\sin^{2}(\theta)\Sigma_{R}(\omega_{+})\right)a_{+}^{\dagger}a_{+}\\
    &+\sin(\theta)\cos(\theta)\left(\left(\Sigma_{R}(\omega_{+},\omega_{-})\right.\right.\\
    &\left.\left.-\Sigma_{L}(\omega_{+},\omega_{-})\right)a_{-}^{\dagger}a_{+}+h.c\right),
    \end{split}
\end{equation}

where we introduced $\Sigma_{\alpha}(\omega)=S_{\alpha}(\omega)+S_{\alpha}(-\omega)$ [$S(\omega)$ is defined in Eq.~\eqref{BathFourier}], and $\Sigma_{\alpha}(\omega,\omega')=S_{\alpha}(-\omega',-\omega)+S_{\alpha}(\omega,\omega')$ [$S(\omega,\omega)$ being defined in Eq.~\eqref{SLS}]. For this model, the off-diagonal contributions of the Lamb-shift take the following simple form

\begin{equation}
\begin{split}
    \Sigma_{\alpha}(\omega_{+},\omega_{-})=&\frac{1}{2\pi}\mathcal{P}\int_{0}^{\infty}d\omega J_{\alpha}(\omega)\left(\frac{1}{\omega_{+}-\omega}+\frac{1}{\omega_{-}-\omega}\right)\\
    &+\frac{J_{\alpha}(\omega_{+})-J_{\alpha}(\omega_{-})}{2i}.
\end{split}
\end{equation}

We remark that it depends only on the spectral density (and thus the coupling strength) of the bath. Therefore we see from Eq.~\eqref{LSOscillator} that if both baths are identically coupled to the oscillators, despite having different temperatures, then the total off-diagonal terms of the Lamb-shift vanish. It implies that the total coupling energy becomes zero. However the coherences of the Lamb-shift associated with each bath are nonzero and contribute to the heat flow exchanged with that bath, as well as to the total entropy production.

For this example, we calculate the total 
entropy production resulting from the dynamics [Eq.~\eqref{total_ep_multibath}] and we compare it to the one obtained with the LNME . For our model the latter is given with the following jump operators and Lamb-shift [see Eq.~\eqref{LNME}].  

\begin{equation}
\begin{split}
A^{(0)}_{\alpha,\overline{\omega}}&=A_{\alpha}(\overline\omega)=a_{\alpha},\\
    H^{(0)}_{LS}&=\Sigma_{L}(\overline{\omega})a^{\dagger}_{L}a_{L}+\Sigma_{R}(\overline{\omega})a^{\dagger}_{R}a_{R}.
\end{split}
\end{equation}

Additionally, we also calculate the naive entropy production that we introduced in the previous example. This allows us to show clearly when an inconsistent thermodynamic behavior of the LNME can appear in the case of multiple baths. We use initial conditions for the system with nonzero coherences in the energy eigenbasis to ensure that $\delta H_S$ plays a nontrivial role.                    %
Based on the previous analysis, we take the following intial state $\rho_{S}(0)\propto \exp\left(-\beta_{L}H_{S}+\mu(a^{\dagger}_{+}a_{-}+a_{+}a^{\dagger}_{-})\right)$ with $\mu\propto\beta_{L}g$. We also take two relatively close inverse temperatures for the baths. Note that when the temperature difference increases, the violations actually vanish, suggesting that the ``classical'' part of the entropy production (i.e the contribution from $\Tr\left(H^{(0)}_{S}D^{(0)}(\rho_{S})\right)$) dominates. The naive entropy production is therefore thermodynamically compatible when coherence is negligible. For the spectral density function of each bath, we again take a Lorentzian $J_{\alpha}(\omega)=c_{\alpha}\omega/(1+(\omega/\Lambda)^{2})$
where the cutoff is assumed to be the same for both baths. In that case we have $\Sigma_{\alpha}(\omega)=c_{\alpha}\Lambda^{3}/2(\Lambda^{2}+\omega^{2})$. We set $\beta_{L}>\beta_{R}$. The methodology for solving the dynamics and calculating the thermodynamic quantities is explained in App.~\ref{appF2}.

The figure \ref{Fig5B1} shows the entropy productions during the dynamics following the considered initial condition for three different settings (GAME, LNME, LNME+naive entropy production definition). In Fig.~\ref{Fig5B1}(a) we have taken $g=5*10^{-3}\overline{\omega}$ and considered a symmetric dissipation strength $c_{L}=c_{R}$. We see that even for such a weak interaction strength, slight violations appear for the naive entropy production (green curve) during the transient regime. Moreover the entropy predicted by GAME (blue curve) is already significantly different from the positive entropy production predicted by the LNME and the zero order expansion of the heat flow compatible with it (red curve). However all the three converge to the same value at the stationary limit. We also show respectively $\Delta\dot\sigma$ and the Spohn entropy production for GAME (respectively the top and bottom insets in Fig.~\ref{Fig5B1}(a)). We see that $\Delta\dot\sigma$ increases and eventually reaches the adiabatic entropy production, while $\dot\sigma_{\text{Sp}}$ is consistently positive and vanishes at the stationary limit.

Let us mention that by decreasing even more the interaction strength $g$, we recovered that the entropy productions for the GAME, the LNME and the naive entropy production eventually match, in what can be called the ``deep local regime''. This is of course expected since the contribution that are $\mathcal{O}(\gamma\Omega)$ become negligible. But with this in mind, we outline the fact that when violations appear, the regime of validity of the LNME starts to be questionable as it miscalculates (in our example it overestimates), compared to GAME, the entropy production during the transient regime. 

In the panel (b) of Fig.~\ref{Fig5B1} we still keep $c_{L}=c_{R}$ but we increased the interaction strength to $g=10^{-2}\overline{\omega}$. As expected, the magnitude of the violations in the naive entropy production increases, along with the discrepancy between the results for GAME and LNME. Again all three values converge at the stationary limit. So far, we always took $c_{L}=c_{R}$ such that the contribution from the off-diagonal terms of the Lamb-shift vanished, as we previously discussed. In Fig.~\ref{Fig5B1}(c) we considered an asymmetric case with $c_{R}=0.9c_{L}$ meaning that the system has a stronger coupling with the left (the coldest) bath, and we kept $g=10^{-2}\overline{\omega}$. We see that the qualitative behavior is similar to the previous case, with however a lower entropy production for the GAME and a discrepancy between the stationary values for the GAME and the LNME (see the inset in Fig.~\ref{Fig5B1}(c)). The naive entropy production predicts the same steady-state value as the GAME, as the contributions from $-i\Tr\left(\delta H_{S}\left[H^{(0)}_{LS},\rho_{S}\right]\right)$ and $\Tr\left(\delta H_{S}D^{(0)}(\rho_{S})\right)$ happen to capture well the corrections predicted by GAME when $t\rightarrow\infty$. 

\begin{figure*}[!t]
  \centering
  \includegraphics[width=0.9\textwidth]{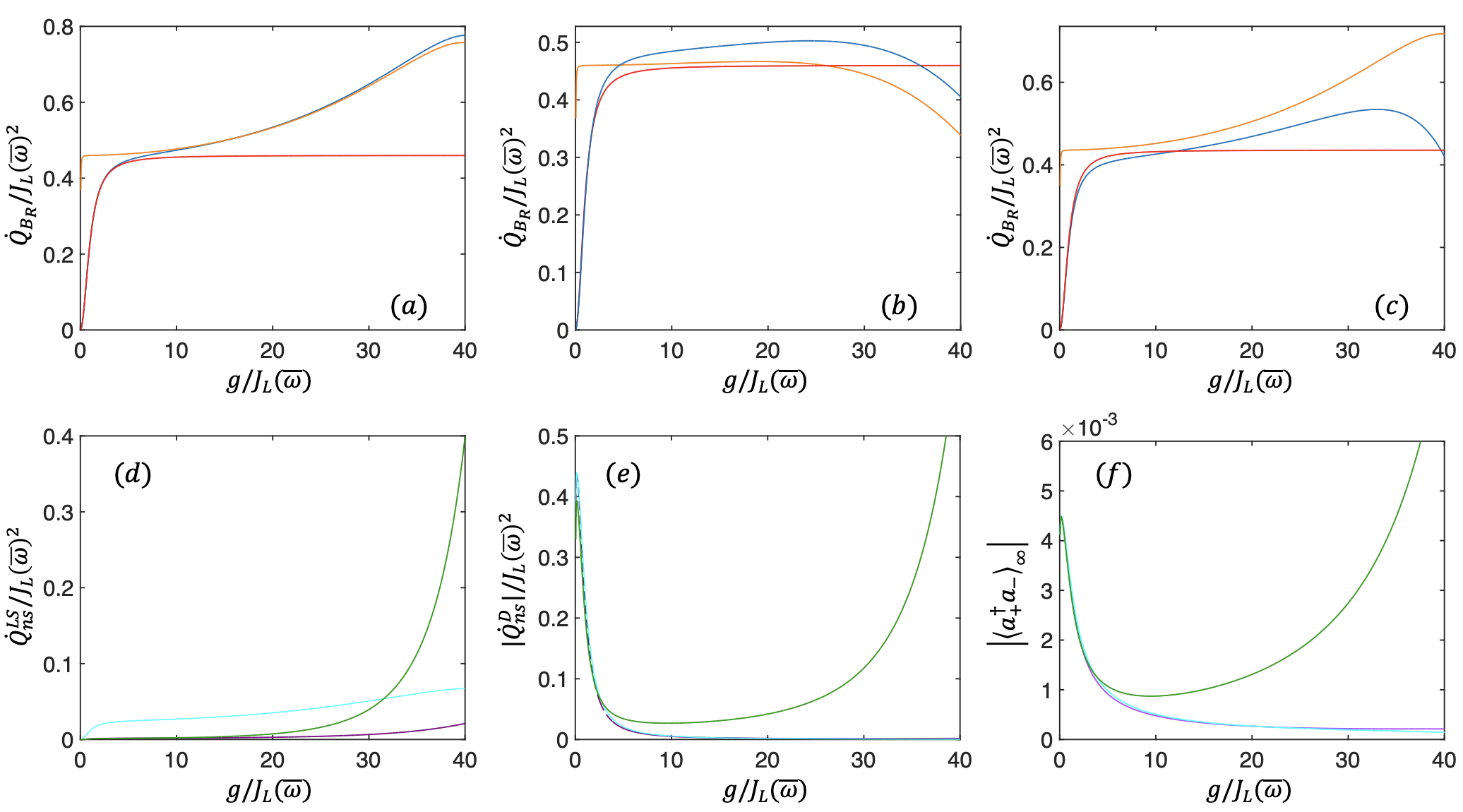}
  \caption{From (a) to (c): stationary heat flux of the right bath as a function of $g$ for three different choices of cutoff and dissipation strength of the baths. The blue curves correspond to the results obtained with GAME, the red ones to the consistent description with the LNME and the orange ones to the heat flux in the full secular regime. The parameters are $\overline{\omega}=50J_{L}(\overline\omega)$, $\delta=10^{-3}\overline{\omega}$, $\beta_{R}\overline{\omega}=1$, $\beta_{L}=\beta_{R}/0.99$. The panel (a) shows the results obtained with $c_{L}\Lambda=0.2\overline{\omega}$ and $c_{R}=c_{L}$. Panel (b) corresponds to $c_{L}\Lambda=0.5\overline\omega$ and $c_{R}=c_{L}$. Panel (c) corresponds to $c_{L}\Lambda=0.2\overline{\omega}$ and $c_{R}=0.9c_{L}$. (d), (e) and (f) are respectively the nonsecular contribution to the heat flux due to the Lamb-shift, the nonsecular contribution due to the dissipator (in absolute value), and the coherence of the system, all three in the stationary state, as functions of $g$. The purple, cyan and green curves correspond respectively to the results obtained with the same settings as the panels (a), (b) and (c).}
  \label{Fig5B2}
\end{figure*}

Our results show that the correction in the entropy production stemming from a systematic treatment of either the GAME or the LNME really matter during the transient regime, while all the approaches seem to give close predictions when the stationary state is reached. To see how far the latter statement is true, we calculated the stationary heat flux of the right bath $\dot Q_{B_{R}}(\infty)$ as a function of $g$. We compare the value obtained with GAME and the consistent thermodynamic description of LNME. Additionally we benchmark our results with the heat flux predicted by the full secular (Lindblad) master equation. This allows us to characterize the deviation of the local regime when $g$ becomes large such that the partial secular approximation is not expected to be valid anymore.    

The obtained results are shown in figure \ref{Fig5B2}(a)-(c), again for the three different settings. The properties of the bath ($c_{\alpha}$ and $\Lambda$) are changed at each panel in order to reveal their influence. For the three cases, we see that we have surprisingly a good match between GAME (blue curves) and LNME (red curves) even for intermediate values of $g$. For instance in panel (a) we have $|\dot Q_{B_R}(\infty)-\dot Q^{(0)}_{B_R}|\leq10^{-3}J_{L}(\overline{\omega})^{2}$ up to $g\approx2.35J_{L}(\overline{\omega})\approx0.047\overline{\omega}$ (and $|\dot Q_{B_R}(\infty)-\dot Q^{(0)}_{B_R}|\leq10^{-2}J_{L}(\overline{\omega})^{2}$ up to $g\approx7.26J_{L}(\overline{\omega})\approx0.15\overline{\omega}$). In panel (a) we picked $c_{L}=c_{R}$ and $c_{L}\Lambda=0.2\overline{\omega}$, and we see that the heat flux predicted by GAME nicely interpolates from LNME to the full secular regime (orange curve). To clarify the comparison between the full secular case and GAME, let us write the expression of the heat flux obtained in the latter case

\begin{equation}
\begin{split}
    \dot Q_{B_R}&=4\Omega\sin(\theta)\cos(\theta)\Im{\Sigma^{*}_{R}(\omega_{+},\omega_{-}))\langle a^{\dagger}_{+}a_{-}\rangle}\\
    &+2\omega_{+}J_{R}(\omega_{+})\sin^{2}(\theta)\left(n_{R}(\omega_{+})-\langle a^{\dagger}_{+}a_{+}\rangle\right)\\
    &+2\omega_{-}J_{R}(\omega_{-})\cos^{2}(\theta)\left(n_{R}(\omega_{-})-\langle a^{\dagger}_{-}a_{-}\rangle\right)\\
    &-2\overline{\omega}\frac{\sqrt{\gamma_{R}(-\omega_{+})\gamma_{R}(-\omega_{-})}}{n_{R}(\overline{\omega})}\Re{\langle a^{\dagger}_{+}a_{-}\rangle}.
\end{split}
\label{RightHeat}
\end{equation}

One can identify three distinct contributions

\begin{equation}
    \begin{split}
        \dot Q_{B_R}&=\dot Q_{s}+\dot Q^{LS}_{ns}+\dot Q^{D}_{ns},\\
        \dot Q_{s}&=2\omega_{+}J_{R}(\omega_{+})\sin^{2}(\theta)\left(n_{R}(\omega_{+})-\langle a^{\dagger}_{+}a_{+}\rangle\right)\\
    &+2\omega_{-}J_{R}(\omega_{-})\cos^{2}(\theta)\left(n_{R}(\omega_{-})-\langle a^{\dagger}_{-}a_{-}\rangle\right),\\
    \dot Q^{LS}_{ns}&=4\Omega\sin(\theta)\cos(\theta)\Im{\Sigma^{*}_{R}(\omega_{+},\omega_{-}))\langle a^{\dagger}_{+}a_{-}\rangle},\\
    \dot Q^{D}_{ns}&=-2\overline{\omega}\frac{\sqrt{\gamma_{R}(-\omega_{+})\gamma_{R}(-\omega_{-})}}{n_{R}(\overline{\omega})}\Re{\langle a^{\dagger}_{+}a_{-}\rangle},
    \end{split}
\end{equation}
where $\dot Q_{s}$ is the contribution from the secular terms (and thus corresponds also to the expression of the heat flux in the full secular regime), $\dot Q^{LS}_{ns}$ is the nonsecular contribution from the Lamb-shift, and $\dot Q^{D}_{ns}$ is the nonsecular contribution from the dissipator. Both $\dot Q^{LS}_{ns}$ and $\dot Q^{D}_{ns}$ correspond to genuine quantum contributions. Let us remark that $\dot Q^{LS}_{ns}$ is at least of order $\mathcal{O}(\gamma\Omega)$ and becomes negligible in the deep local regime, where $\dot Q_{s}$ and $\dot Q^{D}_{ns}$, summed together, lead to the heat flux given by the LNME. Its contribution becomes relevant when $g$ increases. However, at the stationary limit, our perturbative analysis showed that the coherence of the system (quantified here by $\langle a^{+}_{+}a_{-}\rangle$) is given by the correction Hamiltonian $H^{(1)}$. Therefore the nonsecular contributions become negligible when $t\rightarrow\infty$ and the heat flux for GAME is approximately equal to the one in the full secular regime. This is exactly what we see in Fig.~\ref{Fig5B2}(a). To clarify our interpretation, we show in Fig.~\ref{Fig5B2}(d)-(e) respectively $\dot Q^{LS}_{ns}$ and $\dot Q^{D}_{ns}$ (in absolute value), where the purple curves correspond to the results we obtained with the settings of the panel (a). And indeed we can see that the nonsecular contribution from the Lamb-shift remains negligible while the one from the dissipator vanishes when the system leaves the local/PSA regime.

In Fig.~\ref{Fig5B2}(b) we still have $c_{L}=c_{R}$ but we increased the cutoff to $c_{L}\Lambda=0.5\overline{\omega}$. In that case we see a discrepancy between GAME and the full secular regime when $g$ increases. This is explained by the fact that the Lamb-shift contribution is not negligible anymore when $\Lambda$ increases too much. This is illustrated in Fig.~\ref{Fig5B2}(d) where $\dot Q^{LS}_{ns}$ (cyan curve) is significantly larger than in the previous case, however $\dot Q^{D}_{ns}$ is almost identical (Fig.~\ref{Fig5B2}(e)). The discrepancy in the total heat flux increases with $g$ but remains bounded, and the monotonicity is still the same for GAME and the full secular regime. This is because the stationary coherence changes only slightly from panel (a) to (b), as shown in in Fig.~\ref{Fig5B2}(f). In summary, the panel (b) corresponds to a situation where the cutoff frequency of the bath is chosen to be so large that the off-diagonal terms of the Lamb-shift in turn become too large to be fully consistent with the Born-Markov approximation.

Finally in Fig.~\ref{Fig5B2}(c), we reduced the cutoff value back to $c_{L}\Lambda=0.2\overline{\omega}$, but now we picked $c_{R}=0.9c_{L}$, such that the off-diagonal terms of the Lamb-shift (left bath plus right bath contributions) is not zero anymore. As illustrated in Fig.~\ref{Fig5B2}(f), it leads to a drastic change of the stationary coherence (green curve) compared to the two previous cases, when the interaction increases. In particular it grows exponentially when reaching high values of $g$, which is not anymore in agreement with the Born-Markov approximation. We thus have a large discrepancy and a different monotonicity between GAME and the full secular case. This difference is accounted by both $\dot Q^{LS}_{ns}$ and $\dot Q^{D}_{ns}$ (green curves in Fig.~\ref{Fig5B2}(d)-(e)). We are here typically in a case where the choice of bath parameters leads to a strong generation of coherence that is beyond the regime of the validity of our framework. Actually, having off-diagonal Lamb-shift terms that are too large can cause non-Markovian effects during the transient dynamics (revivals) and thus predict a negative entropy production.

\section{Conclusion}\label{sec.V}

In this work, we clarified the thermodynamic description of nonsecular master equations at the average level. Our results show that going beyond the secular regime can fundamentally change the structure of entropy production and energy exchanges. Thermodynamic compatibility is satisfied as long as the energy balance and entropy productions are derived from microscopic definitions, consistently with the approximations made for deriving the master equation. Therefore, 
imposing restrictive conditions such as strict energy conservation or covariance with respect to the bare system Hamiltonian \cite{Dann_21,luo2026trinitymarkovianquantumthermodynamics}, is not necessary and cannot be regarded as universal criteria for recovering thermodynamic consistency in quantum Markovian dynamics.

Beyond the secular regime, the thermodynamics of weakly coupled open quantum systems in fact exhibits several features usually associated to the strong coupling regime. First, the interaction energy has a non-zero contribution in the energy balance. We showed that this contribution is equally shared between the system and bath energy flows and is directly related to the Lamb shift Hamiltonian. Accounting for it yields a consistent first and second law, where the expression of the  heat flow in terms of the system's observables was identified from its microscopic definition.

Another central result beyond the secular regime is that the microscopic entropy production associated with system-bath correlations generally differs from the entropy obtained with the Spohn inequality. The difference contains coherent contributions from the Lamb shift and nonsecular dissipation, which disappear in the global secular regime and are consistently truncated in the local nonsecular limit. For a single thermal bath, these corrections are transient: both entropy production rates and the energy currents vanish in the equilibrium state despite the non-Gibbs form of the reduced system state. We also show that the non-Gibbs form does not imply thermodynamic inconsistency as it does not allow any cyclic work extraction from the single bath. Indeed, even if the stationary state contains extractable work relative to the bare system Hamiltonian, completing a cycle requires switching on and off the system-bath coupling, which costs a greater or equal amount of work.  With multiple baths, the microscopic entropy production accounts for heat currents that persist in the stationary state.

We have then applied and tested the limits of validity of our approach in simple physical models, relevant to highlight the secular to non-secular transition. Despite a focus on nonsecular master equations resulting from the Geometric-Arithmetic approximation, our proposed approach is sufficiently general to suggest that the conclusions and interpretation that we made hold for a wider class of setups. Our examples demonstrate that nonsecular corrections can significantly affect transient thermodynamics. The local approximation can nevertheless remain accurate for stationary quantities over an intermediate range of inter-system coupling strength, while the Geometric-Arithmetic master equation interpolates between the local and global secular regimes. These descriptions resolve the system Bohr frequencies differently, emphasizing the need to match the resolution of thermodynamic observables to that of the chosen master equation.

Our analysis opens the possibility to precisely assess the performance of quantum many body heat engines, including at finite time, and over cycles involving significant variations of the Bohr frequencies. Future works could extend this framework to driven systems and quantum heat engines whose cycles cross the border between secular and nonsecular regime. A complementary direction is a formulation at the trajectory level, investigating how physical time reversal and dual reversal distinguish microscopic and Spohn entropy production and their associated fluctuation relations \cite{Manzano_18,Crooks_2008}. Finally, as it opens the way to further analyze the phenomena specific to the nonsecular regime, our formalism provides an opportunity to gain insights in some features of strong-system bath coupling while avoiding the complexity of non-Markovianity.

\acknowledgements
We thank Camille Lombard-Latune for fruitful feedback and discussions. This work is funded by the European Union. Views and opinions expressed are however those of the authors only and do not necessarily reflect those of the European Union or the European Research Council. Neither the European Union nor the granting authority can be held responsible for them. This work is supported by ERC grant QARNOT, project number 101163469.

\raggedright
\setlength{\bibsep}{0pt}

\bibliography{main}

\appendix
\section{Microscopic derivation of the CGME}\label{appA}
Let us note $\rho_{SB}$ the density matrix that describes the joint state system+bath. We assume the system and the bath to be isolated, so that its dynamics is described by the von Neumann equation. We write it in the interaction picture in order to proceed to a perturbative treatment of the evolution induced by the weak coupling  
\begin{equation}
    \frac{d\tilde\rho_{SB}}{dt}=-i\left[\tilde H_{I}(t),\tilde\rho_{SB}(t)\right],
\end{equation}
with $\tilde{X}(t)=e^{iH_{S}t}e^{iH_{B}t}Xe^{-iH_{S}t}e^{-iH_{B}t}$. We coarse-grain the exact dynamics over $\Delta t$. The Born-Markov approximation implies a separation of timescale between $\tau_{B}$ the time decay of the bath correlation function, and $\tau_{R}$ the relaxation time i.e $\tau_{B}\ll\tau_{R}$, thus we can choose a coarse-graining time satisfying Eq.~\eqref{condition}. We iteratively integrate the von Neumann equation twice between $t$ and $t+\Delta t$, and after tracing out the bath we get the reduced evolution of the system

\begin{align}
\begin{split}
     \tilde\rho_{S}(t+\Delta t)=\tilde\rho_{S}(t)-i\int_{t}^{t+\Delta t}\Tr_{B}\left(\left[\tilde H_{I}(t'),\tilde\rho_{SB}(t')\right]\right)dt'\\
     -\int_{t}^{t+\Delta t}dt'\int_{t}^{t'}\Tr_{B}\left(\left[\tilde H_{I}(t'),\left[H_{I}(t''),\tilde\rho_{SB}(t'')\right]\right]\right)dt''.
\end{split}
\end{align}

Since we integrate over $\Delta t\gg\tau_{B}$, we can assume that the correlations between the system and the bath have a very little contribution in the evolution of $\tilde\rho_{SB}$ between $t$ and $t+\Delta t$ and thus we can replace in the integrals $\rho_{SB}$ by $\rho_{S}\otimes\rho_{B}$. We also assume that the bath operator is centered, i.e $\Tr_{B}\left(B\rho_{B}\right)=0$ (which is common for many cases or can be satisfied by shifting the bath operator without changing the dynamics of the system). With that, the first integral is zero and we get the coarse-grained rate of the variation of the system to be

\begin{widetext}
\begin{equation}
    \frac{\Delta\tilde\rho_{S}}{\Delta t}=-\frac{1}{\Delta t}\int_{t}^{t+\Delta t}dt'\int_{t}^{t'}\Tr_{B}\left(\left[\tilde A(t')\tilde B(t'),\left[\tilde A(t'')\tilde B(t''),\tilde\rho_{S}(t'')\rho_{B}\right]\right]\right)dt''.
\end{equation}
\end{widetext}

We then introduce the time $\tau=t'-t''$ and proceed to a change of variable of the integral. For a fixed $\tau$, we can integrate over $t'$ from $t+\tau$ to $t+\Delta t$, then integrate over $\tau$ from 0 to $\Delta t$. We expand the double commutator and after few lines of algebra we get

\begin{widetext}
    \begin{equation}
    \frac{\Delta\tilde\rho_{S}}{\Delta t}=-\frac{1}{\Delta t}\int_{0}^{\Delta t}d\tau\int_{t+\tau}^{t+\Delta t}g(\tau)\left[\tilde A(t'),\tilde A(t'-\tau)\tilde\rho_{S}(t'-\tau)\right]dt'+h.c,
\end{equation}
\end{widetext}

where $g(\tau)=\Tr_{B}(\tilde B(\tau)B\rho_{B})$ is the bath two-points correlation function. Because $g(\tau)$ is negligible beyond $\tau_{B}$ and $\Delta t\gg\tau_{B}$, we can extend the upper bound of the first integral to $\infty$. We can replace $t+\tau$ by $t$ in the lower bound of the second integral and approximate the state of the system as $\tilde\rho_{S}(t'-\tau)\approx\tilde\rho_{S}(t)$, which gives 

\begin{widetext}
    \begin{equation}
\frac{\Delta\tilde\rho_{S}}{\Delta t}\approx-\frac{1}{\Delta t}\int_{0}^{\infty}d\tau\int_{t}^{t+\Delta t}g(\tau)\left[\tilde A(t'),\tilde A(t'-\tau)\tilde\rho_{S}(t)\right]dt'+h.c.
\end{equation}
\end{widetext}

Let us introduce the eigenstates and eigenenergies of the system Hamiltonian $H_{S}=\sum_{n}\epsilon_{n}\ket{n}\bra{n}$ in order to write the system operators in terms of eigenoperators, $\tilde A(t)=\sum_{\omega}e^{-i\omega t}A(\omega)$ with $A(\omega)=\sum_{\epsilon_{m}-\epsilon_{n}=\omega}\bra{n}A\ket{m}\ket{n}\bra{m}$. We get then

\begin{widetext}
    \begin{equation}
    \frac{\Delta\tilde\rho_{S}}{\Delta t}=-\frac{1}{\Delta t}\sum_{\omega,\omega'}\int_{t}^{t+\Delta t}e^{i(\omega'-\omega)t'}dt'\Gamma(\omega)\left[\ A^{\dagger}(\omega'),A(\omega)\tilde\rho_{S}(t)\right]+h.c,
\end{equation}
\end{widetext}

with $\Gamma(\omega)=\int_{0}^{\infty}e^{i\omega\tau}g(\tau)d\tau$. The integration over $t'$ gives $\int_{t}^{t+\Delta t}e^{i(\omega'-\omega)t'}dt'=\Delta te^{i(\omega'-\omega)t} f_{\Delta t}(\omega,\omega')$ with $f_{\Delta t}(\omega,\omega')=e^{i(\omega'-\omega)\Delta t/2}\text{sinc}((\omega'-\omega)\Delta t/2)$. Then the final step consists of splitting the terms into a unitary part and a non-unitary one. For that we invert the frequencies $\omega$ and $\omega'$ in the sum of the hermitian conjugate term and gather the both sums 

\begin{equation}
\begin{split}
    \frac{\Delta\tilde\rho_{S}}{\Delta t}=&-\sum_{\omega,\omega'}f_{\Delta t}(\omega,\omega')e^{i(\omega'-\omega)t}\left(\Gamma(\omega)\left(\ A^{\dagger}(\omega')A(\omega)\tilde\rho_{S}(t)\right.\right.\\
    &\left.-A(\omega)\tilde\rho_{S}(t)A^{\dagger}(\omega')\right)+\Gamma^{*}(\omega')\left(\tilde\rho_{S}(t)A^{\dagger}(\omega')A(\omega)\right.\\
    &\left.\left.-A(\omega)\tilde\rho_{S}(t)A^{\dagger}(\omega')\right)\right).    
\end{split}
\end{equation}

Finally by introducing the antisymmetric (Eq.~\eqref{SLS}) and symmetric (Eq.~\eqref{gamma2}) parts of $\Gamma(\omega)$ we end up to the final form of the CGME given by Eq.~\eqref{CGME}.

\section{Conservation of energy and detailed derivation of the different contributions}\label{appB}
\subsection{Energy rate of the system}
Here we provide the detailed calculation of the energy rate of the system. For that we consider the dynamics provided by the CGME (Eq.~
\eqref{CGME}) to show that our result is quite general for nonsecular master equations, then we can simply apply the geometric approximation (Eq.~\eqref{geom_approx}) in order to write it for the GAME. The energy rate of the system is given by

\begin{equation}
    \frac{dE_{S}^{\Delta t}}{dt}=-i\Tr_{S}\left(H_{S}\left[H^{\Delta t}_{LS},\rho_{S}\right]\right)+\Tr_{S}\left(H_{S}D^{\Delta t}(\rho_{S})\right).
\end{equation}

Let us first detail the expression of the energy rate due to the Lamb-shift. We have

\begin{align}
    \begin{split}
        &-i\Tr_{S}\left(H_{S}\left[H^{\Delta t}_{LS},\rho_{S}\right]\right)\\
        =&-i\Tr_{S}\left(\rho_{S}\left[H_{S},H^{\Delta t}_{LS}\right]\right)\\
        =&-i\sum_{\omega,\omega'}f_{\Delta t}(\omega,\omega')S(\omega,\omega')\Tr_{S}\left(\rho_{S}\left[H_{S},A^{\dagger}(\omega')A(\omega)\right]\right)\\
        =&-i\sum_{\omega,\omega'}f_{\Delta t}(\omega,\omega')S(\omega,\omega')(\omega'-\omega)\Tr_{S}\left(\rho_{S}A^{\dagger}(\omega')A(\omega)\right),\\
        =&-2\sum_{\omega,\omega'}\omega\Im\left(f_{\Delta t}(\omega,\omega')S(\omega,\omega')\Tr_{S}\left(\rho_{S}A^{\dagger}(\omega')A(\omega)\right)\right).
    \end{split}
\end{align}

We used the cyclic property of the trace to go to the second line and then the fact that $A^{\dagger}(\omega')A(\omega)$ is an eigenoperator of $H_{S}$ at frequency $\omega-\omega'$ by construction. At the last line we used the fact that $f_{\Delta t}(\omega,\omega')$ and $S(\omega,\omega')$ are hermitian with respect to the Bohr frequencies. For the term due to the dissipator we have

\begin{align}
    \begin{split}
        &\Tr_{S}\left(H_{S}D^{\Delta t}(\rho_{S})\right)\\
        =&\sum_{\omega,\omega'}f_{\Delta t}(\omega,\omega')\gamma(\omega,\omega')\left(\Tr_{S}\left(H_{S}A(\omega)\rho_{S}A^{\dagger}(\omega')\right)\right.\\
        &\left.-\frac{1}{2}\Tr_{S}\left(H_{S}\left\{A^{\dagger}(\omega')A(\omega),\rho_{S}\right\}\right)\right)\\
        =&\sum_{\omega,\omega'}f_{\Delta t}(\omega,\omega')\gamma(\omega,\omega')\left(\Tr_{S}\left(\left(-\omega A(\omega)+A(\omega)H_{S}\right)\rho_{S}A^{\dagger}(\omega')\right)\right.\\
        &\left.-\frac{1}{2}\Tr_{S}\left(H_{S}\left\{A^{\dagger}(\omega')A(\omega),\rho_{S}\right\}\right)\right)\\
        =&\sum_{\omega,\omega'}f_{\Delta t}(\omega,\omega')\gamma(\omega,\omega')\left(-\omega\Tr_{S}\left(\rho_{S}A^{\dagger}(\omega)A(\omega)\right)\right.\\
        &\left.+\frac{1}{2}\Tr_{S}\left(\left[A^{\dagger}(\omega')A(\omega),H_{S}\right]\right)\rho_{S}\right)\\
        =&-\frac{1}{2}\sum_{\omega,\omega'}f_{\Delta t}(\omega,\omega')\gamma(\omega,\omega')(\omega+\omega')\Tr_{S}\left(\rho_{S}A^{\dagger}(\omega')A(\omega)\right),\\
        =&-\sum_{\omega,\omega'}\omega \Re\left(f_{\Delta t}(\omega,\omega')\gamma(\omega,\omega')\Tr_{S}\left(\rho_{S}A^{\dagger}(\omega')A(\omega)\right)\right).
    \end{split}
\end{align}

We used the commutation relation $\left[H_{S},A(\omega)\right]=-\omega A(\omega)$ to go to the second line such that an additional $\Tr_{S}\left(H_{S}\rho_{S}A^{\dagger}(\omega')A(\omega)\right)$ term appears, that is then added to the anti-commutator term and simplify into the commutator term at the third equality. Then similarly to the Lamb-shift energy, we used the hermiticity of $f_{\Delta t}(\omega,\omega')$ and $\gamma(\omega,\omega')$ to get the final expression. 

\subsection{Energy rate of the bath}
Here we calculate the energy rate of the bath induced by the coarse-grained evolution of the joint system

\begin{equation}
    \frac{dE_{B}^{\Delta t}}{dt}\approx\Tr\left(H_{B}\left(\frac{\tilde\rho_{SB}(t+\Delta t)-\tilde\rho_{SB}(t)}{\Delta t}\right)\right).
\end{equation}

We can then apply the same approximations as in the derivation of CGME

\begin{widetext}
\begin{equation}
    \frac{dE_{B}^{\Delta t}}{dt}\approx-\frac{1}{\Delta t}\int_{0}^{\Delta t}d\tau\int_{t+\tau}^{t+\Delta t}\Tr\left(H_{B}\left[\tilde A(t')\tilde B(t'),\left[\tilde A(t'-\tau)\tilde B(t'-\tau),\tilde\rho_{S}(t'-\tau)\rho_{B}\right]\right]\right)dt'.
\end{equation}    
\end{widetext}

We can then expand the double commutators and use the fact that $H_{B}$ commutes with $\rho_{B}$. After a few lines of algebras and using the Markov approximation and $\Delta t\gg\tau_{B}$, we get

\begin{widetext}
\begin{equation}\label{dEB}
    \frac{dE_{B}^{\Delta t}}{dt}\approx\frac{1}{\Delta t}\int_{0}^{\infty}d\tau\int_{t}^{t+\Delta t}\Tr_{S}\left(\tilde\rho_{S}(t)A(t')A(t'-\tau)\right)\Tr_{B}\left(\left[\tilde B(\tau),H_{B}\right]B\rho_{B}\right)dt'+c.c.
\end{equation}    
\end{widetext}

Then we can write the system operator in terms of eigenoperators to get

\begin{widetext}
\begin{equation}
    \frac{dE_{B}^{\Delta t}}{dt}=\frac{1}{\Delta t}\sum_{\omega,\omega'}\int_{t}^{t+\Delta t}e^{i(\omega'-\omega)t'}dt'E(\omega)\Tr_{S}\left(\rho_{S}A^{\dagger}(\omega')A(\omega)\right)+c.c,
\end{equation}     
\end{widetext}

where $E(\omega)=\int_{0}^{\infty}e^{i\omega\tau}\Tr_{B}\left(\left[\tilde B(\tau),H_{B}\right]B\rho_{B}\right)d\tau$. Like for the CGME the integration over $t'$ leads to $\Delta te^{i(\omega'-\omega)t} f_{\Delta t}(\omega,\omega')$. Now let us work out the expression of $E(\omega)$. For that we will introduce the eigenoperators of $B$ to rewrite $E(\omega)$ as

\begin{align}
    \begin{split}
    E(\omega)&=\int_{0}^{\infty}e^{i\omega\tau}\Tr_{B}\left(\left[\tilde B(\tau),H_{B}\right]B\rho_{B}\right)d\tau\\
    &=\sum_{\nu}\int_{0}^{\infty}e^{i(\omega-\nu)\tau}\Tr_{B}\left(\left[ B(\nu),H_{B}\right]B\rho_{B}\right)d\tau\\
    &=\sum_{\nu}\int_{0}^{\infty}e^{i(\omega-\nu)\tau}\nu\Tr_{B}\left( B(\nu)B\rho_{B}\right)d\tau\\
    &=i\int_{0}^{\infty}e^{i\omega\tau}\frac{d}{d\tau}\left(\sum_{\nu}e^{-i\nu\tau}\Tr_{B}\left( B(\nu)B\rho_{B}\right)\right)d\tau\\
    &=i\int_{0}^{\infty}e^{i\omega\tau}\frac{dg}{d\tau}d\tau\\
    &=i\left(\left[e^{i\omega\tau}g(\tau)\right]^{\infty}_{0}-i\omega\int_{0}^{\infty}e^{i\omega\tau}g(\tau)d\tau\right)\\
    &=-ig(0)+\omega\Gamma(\omega).
    \end{split}
\end{align}

Let us remark that the bath can have a continuous spectrum. In that case we can introduce an integral over $\nu$ with a spectral density function, but the final result does not change. At the last line we used the fact that $\lim_{\tau\rightarrow\infty}g(\tau)=0$ (Markov approximation). Now we can gather the first sum and its complex conjugate as

\begin{widetext}
\begin{equation}
\begin{split}
    \frac{dE_{B}^{\Delta t}}{dt}&=\sum_{\omega,\omega'}e^{i(\omega'-\omega)t}f_{\Delta t}(\omega,\omega')\left(\omega\Gamma(\omega)+\omega'\Gamma^{*}(\omega')\right)\Tr_{S}\left(\tilde\rho_{S}A^{\dagger}(\omega')A(\omega)\right)\\
    &=\sum_{\omega,\omega'}e^{i(\omega'-\omega)t}f_{\Delta t}(\omega,\omega')\left(\left(\frac{\omega+\omega'}{2}+\frac{\omega-\omega'}{2}\right)\Gamma(\omega)+\left(\frac{\omega+\omega'}{2}+\frac{\omega'-\omega}{2}\right)\Gamma^{*}(\omega')\right)\Tr_{S}\left(\tilde\rho_{S}A^{\dagger}(\omega')A(\omega)\right)\\
    &=\sum_{\omega,\omega'}e^{i(\omega'-\omega)t}f_{\Delta t}(\omega,\omega')\left(\frac{\omega+\omega'}{2}\gamma(\omega,\omega')+i(\omega-\omega')S(\omega,\omega')\right)\Tr_{S}\left(\tilde\rho_{S}A^{\dagger}(\omega')A(\omega)\right).
\end{split}
\end{equation}    
\end{widetext}

The term $-ig(0)$ gets cancel with its complex conjugate since $g(0)$ is real. Finally, we can clearly see that the energy rate of the bath is related to the energy rate of the system as 

\begin{equation}
    \frac{dE_{B}^{\Delta t}}{dt}=-i\Tr_{S}\left(H_{S}\left[H^{\Delta t}_{LS},\rho_{S}\right]\right)-\Tr_{S}\left(H_{S}D^{\Delta t}(\rho_{S})\right).
\end{equation}

\subsection{Energy rate of the coupling term}

Let us start from the exact expression of the energy rate

\begin{equation}
    \frac{dE_{I}}{dt}=\Tr\left(\frac{d\tilde H_{I}}{dt}\tilde\rho_{SB}(t)\right)+\Tr\left(\tilde H_{I}(t)\frac{d\tilde\rho_{SB}(t)}{dt}\right).
\end{equation}

Then the coarse grained rate is given by
\begin{widetext}
    \begin{equation}
    \begin{split}
         \frac{dE^{\Delta t}_{I}}{dt}&=\frac{1}{\Delta t}\int_{t}^{t+\Delta t}\frac{dE_{I}}{dt'}dt'\\
         &=\frac{1}{\Delta t}\int_{t}^{t+\Delta t}\Tr\left(\frac{d\tilde H_{I}}{dt'}\tilde\rho_{SB}(t')\right)+\Tr\left(\tilde H_{I}(t')\frac{d\tilde\rho_{SB}}{dt'}\right)dt'\\
         &=\frac{1}{\Delta t}\left(i\Tr\left(\int_{t}^{t+\Delta t}\left[H_{S}+H_{B},\tilde H_{I}(t')\right]\tilde\rho_{SB}(t')dt'\right)-i\underbrace{\Tr\left(\int_{t}^{t+\Delta t}\tilde H_{I}(t')\left[\tilde H_{I}(t'),\tilde\rho_{SB}(t')\right]dt'\right)}_\textrm{=0}\right)\\
         &=\frac{i}{\Delta t}\left(\Tr\left(\int_{t}^{t+\Delta t}\left[H_{S}+H_{B},\tilde H_{I}(t')\right]\tilde\rho_{SB}(t)dt'\right)-i\Tr\left(\int_{t}^{t+\Delta t}\left[H_{S}+H_{B},\tilde H_{I}(t')\right]dt'\int_{t}^{t'}\left[\tilde H_{I}(t''),\tilde\rho_{SB}(t'')\right]dt''\right)\right).
    \end{split}
\end{equation}
\end{widetext}

By using the approximation $\tilde\rho_{SB}(t)\approx\tilde\rho_{S}(t)\rho_{B}$ and the assumption $\Tr_{B}(\tilde B(t)\rho_{B})=0$ we can deduce that $\Tr\left(\left[H_{S},\tilde H_{I}(t')\right]\tilde\rho_{SB}(t)\right)=0$. In addition since $H_{B}$ commutes with $\rho_{B}$ then we also end up with $\Tr\left(\left[H_{B},\tilde H_{I}(t')\right]\tilde\rho_{SB}(t)\right)=0$ and thus the first integral is null. Let us work out the second term, we introduce the time $\tau=t'-t''$ and use the Born-Markov approximation to get

\begin{widetext}
    \begin{equation}
    \begin{split}
        \frac{dE^{\Delta t}_{I}}{dt}&\approx\frac{1}{\Delta t}\int_{0}^{\infty}d\tau\int_{t}^{t+\Delta t}\Tr\left(\left[H_{S}+H_{B},\tilde A(t')\tilde B(t')\right]\left[\tilde A(t'-\tau)\tilde B(t'-\tau),\tilde\rho_{S}(t)\rho_{B}\right]\right)dt'.
    \end{split}
\end{equation}
\end{widetext}

We see that the energy rate contains two contribution, one involving the system Hamiltonian and the other the bath Hamiltonian. We will now show they are respectively related to the energy of the system and the bath. Let us focus first on the term involving $H_{B}$. After expanding the two commutators we can rewrite it as

\begin{widetext}
\begin{equation}
    \begin{split}
        &\frac{1}{\Delta t}\int_{0}^{\infty}d\tau\int_{t}^{t+\Delta t}\Tr\left(\left[H_{B},\tilde A(t')\tilde B(t')\right]\left[\tilde A(t'-\tau)\tilde B(t'-\tau),\tilde\rho_{S}(t)\rho_{B}\right]\right)dt'\\
        &=\frac{1}{\Delta t}\int_{t}^{t+\Delta t}dt'\int_{0}^{\infty}\Tr_{S}\left(\tilde\rho_{S}(t)\tilde A(t')\tilde A(t'-\tau)\right)\Tr_{B}\left(\left[H_{B},\tilde B(\tau)\right]B\rho_{B}\right)+c.c.
    \end{split}
\end{equation}
\end{widetext}

We can compare to Eq.~\eqref{dEB} and see that we recovered exactly the opposite of the energy rate of the bath. Finally let us work out the term involving the system Hamiltonian. Again by expanding the two commutators, we get

\begin{widetext}
    \begin{equation}
        \begin{split}
            &\frac{1}{\Delta t}\int_{0}^{\infty}d\tau\int_{t}^{t+\Delta t}\Tr\left(\left[H_{S},\tilde A(t')\tilde B(t')\right]\left[\tilde A(t'-\tau)\tilde B(t'-\tau),\tilde\rho_{S}(t)\rho_{B}\right]\right)dt'\\
            &=\frac{1}{\Delta t}\int_{t}^{t+\Delta t}dt'\int_{0}^{\infty}g(\tau)\Tr_{S}\left(\left[H_{S},\tilde A(t')\right]\tilde A(t'-\tau)\tilde\rho_{S}\right)d\tau+c.c\\
            &=\frac{1}{\Delta t}\sum_{\omega,\omega'}\int_{t}^{t+\Delta t}e^{i(\omega'-\omega)t'}dt'\Gamma(\omega)\Tr_{S}\left(\left[H_{S}, A^{\dagger}(\omega')\right] A(\omega)\tilde\rho_{S}\right)d\tau+c.c\\
            &=\sum_{\omega,\omega'}e^{i(\omega'-\omega)t}f_{\Delta t}(\omega,\omega')\left(\omega'\Gamma(\omega)+\omega\Gamma^{*}(\omega')\right)\Tr_{S}\left(A^{\dagger}(\omega') A(\omega)\tilde\rho_{S}\right)\\
            &=\sum_{\omega,\omega'}e^{i(\omega'-\omega)t}f_{\Delta t}(\omega,\omega')\left(\frac{\omega'+\omega}{2}\gamma(\omega,\omega')+i(\omega'-\omega)S(\omega,\omega')\right)\Tr_{S}\left(A^{\dagger}(\omega') A(\omega)\tilde\rho_{S}\right)\\
            &=-\frac{dE^{\Delta t}_{S}}{dt}.
        \end{split}
    \end{equation}
\end{widetext}

So we proved the conservation of energy for the CGME, therefore we can derive it for any specific master equation (Redfield, Lindblad, LNME, GAME) after applying the appropriate secular/partial secular approximation.

\section{Perturbative treatment of the steady state of the GAME}\label{appC}

\subsection{Determination of the dominant term \texorpdfstring{$\rho_{\infty}^{(0)}$}{rho-infty(0)}}

As shown in \ref{sec.IIIB}, the dominant term is diagonal in the basis of $H_{S}$ and we can write it as $\rho_{\infty}^{(0)}=\sum_{n}p_{n}^{(0)}\ket{n}\bra{n}$. We will now exploit its simple structure in order to get its expression with the second relation of Eq.~\eqref{exp_st}. We simply project it in the diagonal elements of $H_{S}$ (therefore the commutator terms vanish)
\begin{equation}
    \bra{n}D^{G}(\rho_{_\infty}^{(0)})\ket{n}=0.
\end{equation}
Let's calculate the different terms inside the double sum of the dissipator (Eq.~\eqref{GAME})
\begin{equation}
    \begin{split}
        &\bra{n}A(\omega)\rho_{\infty}^{(0)}A^{\dagger}(\omega')\ket{n}\\
        &=\sum_{m}\bra{n}A(\omega)p_{m}^{(0)}\ket{m}\bra{m}A^{\dagger}(\omega')\ket{n}\\
        &=\sum_{m}p_{m}^{(0)}\bra{n}A(\omega)\ket{m}\bra{m}A^{\dagger}(\omega')\ket{n}\\
        &=\sum_{m}\delta_{\omega,\omega'}\delta_{\omega,\epsilon_{m}-\epsilon_{n}}p_{m}^{(0)}\bra{n}A(\omega)\ket{m}\bra{m}A^{\dagger}(\omega')\ket{n},
    \end{split}
\end{equation}
where $\epsilon_{n}$ is the eigenenergy and we used the property $\bra{n}A(\omega)\ket{m}\neq0$ if $\epsilon_{m}-\epsilon_{n}=\omega$ . Now the second term in the sum gives us
\begin{equation}
\begin{split}
    &\frac{1}{2}\bra{n}\left\{A^{\dagger}(\omega')A(\omega),\rho_{\infty}^{(0)}\right\}\ket{n}
    \\&=\frac{1}{2}\sum_{m}\bra{n}A^{\dagger}(\omega')A(\omega)p_{m}^{0}\ket{m}\bra{m}\ket{n}\\
    &+\bra{n}p_{m}^{0}\ket{m}\bra{m}A^{\dagger}(\omega')A(\omega)\ket{n}\\
    &=p_{n}^{(0)}\bra{n}A^{\dagger}(\omega')A(\omega)\ket{n}\\
    &=p_{n}^{(0)}\sum_{m}\bra{n}A^{\dagger}(\omega')\ket{m}\bra{m}A(\omega)\ket{n}\\
    &=p_{n}^{(0)}\sum_{m}\delta_{\omega,\omega'}\delta_{\omega,\epsilon_{n}-\epsilon_{m}}\bra{n}A^{\dagger}(\omega')\ket{m}\bra{m}A(\omega)\ket{n}
\end{split}
\end{equation}
Therefore by using the Kronecker $\delta_{\omega,\omega'}$ we get 
\begin{equation}
\begin{split}
    &\bra{n}D^{G}(\rho_{_\infty}^{(0)})\ket{n}\\
    &=\sum_{\omega}\sum_{m}\gamma(\omega)\left(\delta_{\omega,\epsilon_{m}-\epsilon_{n}}p_{m}^{(0)}\bra{n}A(\omega)\ket{m}\bra{m}A^{\dagger}(\omega)\ket{n}\right.\\
    &\left.+p_{n}^{(0)}\delta_{\omega,\epsilon_{n}-\epsilon_{m}}\bra{n}A^{\dagger}(\omega)\ket{m}\bra{m}A(\omega)\ket{n}\right)=0.
\end{split}
\end{equation}

We will now modify the second term in the sum by changing the frequency $\omega$ to $-\omega$ and use that $A^{\dagger}(\omega)=A(-\omega)$ so that we can factorize all terms with the same matrix elements

\begin{equation}
\begin{split} \sum_{\omega}\sum_{m}&|\bra{n}A(\omega)\ket{m}|^{2}\delta_{\omega,\epsilon_{m}-\epsilon_{n}}\left(\gamma(\omega)p_{m}^{(0)}\right.\\
    &\left.-\gamma(-\omega)p_{n}^{(0)}\right)=0.
\end{split}
\end{equation}

We can thus deduce that the terms inside the brackets have to be equal to zero. Now we use the KMS condition $\gamma(-\omega)=e^{-\beta\omega}\gamma(\omega)$ and the last selection rule given by $\delta_{\omega,\epsilon_{m}-\epsilon_{n}}$ to finally get 

\begin{equation}
    p_{m}^{(0)}=e^{-\beta(\epsilon_{m}-\epsilon_{n})}p_{n}^{(0)}.
\end{equation}
From this expression we can simply deduce that the dominant term of the steady state is given by the Gibbs state with the respect to the system Hamiltonian  i.e $\rho_{\infty}^{(0)}=e^{-\beta H_{S}}/Z_{S}$.

\subsection{Next order term of the steady state and correction Hamiltonian}

Here we show how the correction Hamiltonian $H^{(1)}$ is related to $\rho_{\infty}^{(1)}$ and derive its expression. First we do a first order Dyson expansion of the effective Gibbs state by treating the inverse temperature as an imaginary time $\left(e^{-\beta H_{S}}=U_{S}(-i\beta)\ \text{with}\ U_{S}(t)=e^{-i H_{S}t}\right)$

\begin{equation}
\begin{split}
    e^{-\beta H_\text{eff}}&=e^{-\beta H_{S}}\mathcal{T}e^{\gamma\int_{0}^{\beta}\tilde{H}_{1}(-iu)du}\\
    &\approx e^{-\beta H_{S}}\left(\mathbb{I}-\gamma\int_{0}^{\beta}e^{u H_{S}}H^{(1)}e^{-uH_{S}}du\right)
\end{split}
\end{equation}

The integral inside the bracket can actually be simply calculated by using the well known commutation relation between the eigenoperators and the Gibbs state
\begin{equation}
        A(\omega)e^{-\beta H_{S}}=e^{-\beta\omega}e^{-\beta H_{S}}A(\omega).
\end{equation}
We get then

\begin{equation}
    \begin{split}
        &\int_{0}^{\beta}e^{u H_{S}}H^{(1)}e^{-uH_{S}}du\\
        &=\sum_{\overline\omega}\sum_{\omega,\omega'\in\mathcal{L}(\overline{\omega})}H^{(1)}(\omega,\omega')\int_{0}^{\beta}e^{u H_{S}}A^{\dagger}(\omega')A(\omega)e^{-uH_{S}}du\\
        &=\sum_{\overline\omega}\sum_{\omega,\omega'\in\mathcal{L}(\overline{\omega})}H^{(1)}(\omega,\omega')A^{\dagger}(\omega')A(\omega)\int_{0}^{\beta}e^{u(\omega'-\omega)}du\\
        &=\sum_{\overline\omega}\sum_{\omega,\omega'\in\mathcal{L}(\overline{\omega})}H^{(1)}(\omega,\omega')\alpha(\omega'-\omega) A^{\dagger}(\omega')A(\omega),
    \end{split}
\end{equation}

with $\alpha(x)=(e^{\beta x}-1)/x$. Using the renormalization condition $\Tr(\rho_{\infty}^{(1)})=0$, we get the expression of $\rho_{\infty}^{(1)}$ in terms of the elements of $H^{(1)}$ (Eq.~\eqref{rhoinf1}). Finally we determine the elements $H^{(1)}(\omega,\omega')$, by using the second relation in Eq.~\eqref{exp_st}. Let us calculate the different terms:

\begin{widetext}
\begin{equation}
\begin{split}
&-i[H_{S},\rho_{\infty}^{(1)}]\\
&=i\left[H_{S},\sum_{\overline\omega}\sum_{\omega,\omega'\in\mathcal{L}(\overline{\omega})}H^{(1)}(\omega,\omega')\alpha(\omega'-\omega)\rho_{\infty}^{(0)} A^{\dagger}(\omega')A(\omega)\right]\\
        &=i\sum_{\overline\omega}\sum_{\omega,\omega'\in\mathcal{L}(\overline{\omega})}H^{(1)}(\omega,\omega')\alpha(\omega'-\omega)(\omega'-\omega)\rho_{\infty}^{(0)} A^{\dagger}(\omega')A(\omega)\\
        &=i\sum_{\overline\omega}\sum_{\omega,\omega'\in\mathcal{L}(\overline{\omega})}H^{(1)}(\omega,\omega')(e^{\beta(\omega'-\omega)}-1)\rho_{\infty}^{(0)} A^{\dagger}(\omega')A(\omega),      
\end{split}
\end{equation}    
\end{widetext}

where we used the fact that $\rho_{\infty}^{(0)}$ commutes with $H_{S}$ and $\left[{H_{S},A^{\dagger}(\omega')A(\omega)}\right]=(\omega'-\omega)A^{\dagger}(\omega')A(\omega)$ to go to the second line. The term involving the Lamb-shift becomes

\begin{equation}
    \begin{split}
        &-i[H_{LS},\rho_{\infty}^{(0)}]\\
        &=-i\left[\sum_{\overline\omega}\sum_{\omega,\omega'\in\mathcal{L}(\overline{\omega})}S_{ij}(\omega,\omega')A^{\dagger}(\omega')A(\omega),\rho_{\infty}^{(0)}\right]\\
        &=-i\sum_{\overline\omega}\sum_{\omega,\omega'\in\mathcal{L}(\overline{\omega})}S_{ij}(\omega,\omega')(e^{\beta(\omega'-\omega)}-1)\rho_{\infty}^{(0)}A^{\dagger}(\omega')A(\omega).
    \end{split}
\end{equation}

Finally the dissipator
\begin{equation}
    \begin{split}
        &D^{G}(\rho_{\infty}^{(0)})\\
        &=\sum_{\overline\omega}\sum_{\omega,\omega'\in\mathcal{L}(\overline{\omega})}\sqrt{\gamma(\omega)\gamma(\omega')}\left(A(\omega)\rho_{\infty}^{(0)} A^{\dagger}(\omega')\right.\\
        &\left.-\frac{1}{2}\left\{A^{\dagger}(\omega')A(\omega),\rho_{\infty}^{(0)}\right\}\right)\\
        &=\sum_{\overline\omega}\sum_{\omega,\omega'\in\mathcal{L}(\overline{\omega})}\sqrt{\gamma(\omega)\gamma(\omega')}\left(e^{-\beta\omega}\rho_{\infty}^{(0)} A(\omega)A^{\dagger}(\omega')\right.\\
        &\left.-\frac{1}{2}\left(1+e^{\beta(\omega'-\omega)}\right)\rho_{\infty}^{(0)}A^{\dagger}(\omega')A(\omega)\right)\\
        &=\sum_{\overline\omega}\sum_{\omega,\omega'\in\mathcal{L}(\overline{\omega})}\left(\sqrt{\gamma(-\omega')\gamma(-\omega)}e^{\beta\omega'}\right.\\
        &\left.-\frac{\sqrt{\gamma(\omega)\gamma(\omega')}}{2}\left(1+e^{\beta(\omega'-\omega)}\right)\right)\rho_{\infty}^{(0)}A^{\dagger}(\omega')A(\omega)\\
        &=\sum_{\overline\omega}\sum_{\omega,\omega'\in\mathcal{L}(\overline{\omega})}\sqrt{\gamma(\omega')\gamma(\omega)}\left(e^{\beta(\omega'-\omega)/2}\right.\\
        &\left.-\left(1+e^{\beta(\omega'-\omega)}\right)/2\right)\rho_{\infty}^{(0)}A^{\dagger}(\omega')A(\omega).
    \end{split}
\end{equation}

To go from the second line to the third one, we manipulated the first term in the sum by switching the frequencies $\omega\rightarrow-\omega'$ and $\omega'\rightarrow-\omega$, and used $A(-\omega)=A^{\dagger}(\omega)$. We then used the KMS condition to get the last line. Now all the different terms can be factorized with respect to $\rho_{\infty}^{(0)}A^{\dagger}(\omega')A(\omega)$ and since it is equal to zero then it means that each terms in the sum has to be equal zero i.e

\begin{equation}
    \begin{split}
        &iH^{(1)}(\omega,\omega')(e^{\beta(\omega'-\omega)}-1)-iS_{ij}(\omega,\omega')(e^{\beta(\omega'-\omega)}-1)\\
        &+\sqrt{\gamma(\omega')\gamma(\omega)}\left(e^{\beta(\omega'-\omega)/2}-\frac{\left(1+e^{\beta(\omega'-\omega)}\right)}{2}\right)=0\\
        &\implies H^{(1)}(\omega,\omega')=S(\omega,\omega')+\frac{i\sqrt{\gamma(\omega')\gamma(\omega)}}{e^{\beta(\omega'-\omega)}-1}\left(e^{\beta(\omega'-\omega)/2}\right.\\
        &-\left.\frac{\left(1+e^{\beta(\omega'-\omega)}\right)}{2}\right).
    \end{split}
\end{equation}

We can do the identification $H^{(1)}=H_{LS}+K$ and the expression of K can be written in a more compact way
\begin{equation}\label{K2}
    \begin{split}
        K(\omega,\omega')&=\frac{i\sqrt{\gamma(\omega')\gamma(\omega)}}{e^{\beta(\omega'-\omega)}-1}\left(e^{\beta(\omega'-\omega)/2}-\frac{\left(1+e^{\beta(\omega'-\omega)}\right)}{2}\right)\\
        &=\frac{-i\sqrt{\gamma(\omega')\gamma(\omega)}}{2(e^{\beta(\omega'-\omega)}-1)}\left(e^{\beta(\omega'-\omega)/2}-1\right)^2\\
        &=\frac{-i\sqrt{\gamma(\omega')\gamma(\omega)}}{2(e^{\beta(\omega'-\omega)/2}+1)}\left(e^{\beta(\omega'-\omega)/2}-1\right)\\
        &=\frac{i\sqrt{\gamma(\omega')\gamma(\omega)}\tanh(\beta(\omega-\omega')/4)}{2}
    \end{split}
\end{equation}

It is easy to check that $K$ is hermitian. 

\subsection{Spohn inequality for the GAME}

Now we can derive the expression for the Spohn entropy production (we made appear explicitly the scaling $\gamma$ for clarity)

\begin{equation}
    \begin{split}
        &\dot{\sigma}_{Sp}(t)=-\frac{dS(\rho(t)||\rho_{\infty})}{dt}\\
        &=\frac{dS}{dt}+\Tr\left(\frac{d\rho_{S}}{dt}\log(\rho_{\infty})\right)\\
        &=\frac{dS}{dt}-\beta\Tr\left(\left(L^{(0)}(\rho_{S})+\gamma L^{(1)}(\rho_{S})\right)\left(H_{S}+\gamma H^{(1)}\right)\right)\\
        &\approx\frac{dS}{dt}-\gamma\beta\left(-i\Tr\left([H_{S},\rho_{S}]H^{(1)}\right)-i\Tr\left([H_{LS},\rho_{S}]H_{S}\right)\right.\\
        &\quad\left.+\Tr\left(H_{S}D^{G}(\rho_{S})\right)\right)+\mathcal{O}(\gamma^{2})\\
        &=\frac{dS}{dt}-\gamma\beta\left(-i\Tr\left([H_{LS}-H^{(1)},\rho_{S}]H_{S}\right)\right.\\
        &\quad\left.+\Tr\left(H_{S}D^{G}(\rho_{S})\right)\right)\\
        &=\frac{dS}{dt}-\gamma\beta\left(i\Tr\left([K,\rho_{S}]H_{S}\right)+\Tr\left(H_{S}D^{G}(\rho_{S})\right)\right).\\   
    \end{split}
\end{equation}

\section{Bound for the ergotropy of the GAME steady state}\label{appD}

The integrated version of the second law (Eq.~\eqref{d:sigmatot}) leads us to

\begin{equation}
    \Delta S+\beta\Delta E_{B}\geq0.
\end{equation}

Then we use the conservation of energy $\Delta E_{B}=-\Delta E_{S}-\Delta E_{I}$ and introduce the non-equilibrium free energy $F_{\beta}(\rho)$ to get

\begin{equation}
    \Delta F_{\beta}\leq-\Delta E_{I}
\end{equation}

Then we can relate the variation of the free energy to the ergotropy by introducing the free energy of the passive state $\rho_{P}$ as

\begin{equation}
    \begin{split}
        \Delta F_{\beta}&=(F_{\beta}(\rho_{\infty})-F_{\beta}(\rho_{P}))+F_{\beta}(\rho_{P})-F_{\beta}(\rho_{S}(0))\\
        &=R(\rho_{\infty})+F_{\beta}(\rho_{P})-F_{\beta}(\rho_{S}(0))\\
        &\geq R(\rho_{\infty})+F_{\beta}(\rho_{\beta})-F_{\beta}(\rho_{S}(0)).
    \end{split}
\end{equation}

To go from the first to the second line, we used the fact that $S(\rho_{\infty})=S(\rho_{P})$, and to get the inequality we used the that $\rho_{\beta}$ is the state that minimizes the free energy. In the end we deduce

\begin{equation}
    R(\rho_{\infty})\leq-\Delta E_{I}+F_{\beta}(\rho_{S}(0))-F_{\beta}(\rho_{\beta}).
\end{equation}

\section{Steady state in the presence of multiple baths}\label{appE}

Like in the case of one thermal bath, we get the steady state to be expanded as $\rho_{\infty}\approx\rho^{(0)}_{\infty}+\gamma\rho^{(1)}_{\infty}$ with $\rho^{(0)}_{\infty}=\sum_{n}p^{(0)}_{n}\ket{n}\bra{n}$. After projecting the dissipator on an eigenstate of $H_{S}$, we get 

\begin{equation}
\begin{split}
    &\sum_{\alpha}\bra{n}D_{\alpha}^{G}(\rho_{_\infty}^{(0)})\ket{n}\\
&=\sum_{\alpha}\sum_{\omega}\sum_{m}|\bra{n}A_{\alpha}(\omega)\ket{m}|^{2}\delta_{\omega,\epsilon_{m}-\epsilon_{n}}\left(\gamma_{\alpha}(\omega)p_{m}^{(0)}\right.\\
    &\left.-\gamma_{\alpha}(-\omega)p_{n}^{(0)}\right)=0.
\end{split}
\end{equation}

We can then use the KMS relation for each bath $\gamma_{\alpha}(-\omega)=e^{-\beta_{\alpha}\omega}\gamma_{\alpha}(\omega)$, however we do not get in the end a simple relation between $p_{n}^{(0)}$ and $p_{m}^{(0)}$. In order to obtain a proportional relation like in the case of a single bath, we can introduce an effective inverse temperature as follow
\begin{equation}
\begin{split}
    &\sum_{\alpha}|\bra{n}A_{\alpha}(\omega)\ket{m}|^{2}e^{-\beta_{\alpha}\omega}\gamma_{\alpha}(\omega)\\
    &=e^{-\tilde\beta(\omega)\omega}\sum_{\alpha}|\bra{n}A_{\alpha}(\omega)\ket{m}|^{2}\gamma_{\alpha}(\omega),\\
    &\implies \tilde\beta(\omega)=-\frac{1}{\omega}\log\left(\frac{\sum_{\alpha}|\bra{n}A(\omega)\ket{m}|^{2}e^{-\beta_{\alpha}\omega}\gamma_{\alpha}(\omega)}{\sum_{\alpha}|\bra{n}A(\omega)\ket{m}|^{2}\gamma_{\alpha}(\omega)}\right).
\end{split}
\end{equation}

Nevertheless, the KMS relation ensures that the effective inverse temperature is an even function of frequency and is always positive. Then we can use the selection rule and deduce that the occupation populations satisfy the following effective detailed balance relation

\begin{equation}
    p_{m}^{(0)}=e^{-\tilde\beta(\epsilon_{m}-\epsilon_{n})(\epsilon_{m}-\epsilon_{n})}p_{n}^{(0)}.
\end{equation}

Next we can determine the form of $\rho^{(1)}_{\infty}$. Inspired by the derivation for the single bath case, we assume that it can be written as  

\begin{equation}
    \begin{split} \rho_{\infty}^{(1)}=&\sum_{\alpha}\sum_{\overline{\omega}}\sum_{\omega,\omega'\in\mathcal{L(\overline{\omega})}}R_{\alpha}(\omega,\omega')\rho_{\infty}^{(0)}\left(A^{\dagger}_{\alpha}(\omega')A_{\alpha}(\omega)\right.\\
        &\left.-\Tr_{S}\left(\rho_{\infty}^{(0)}A^{\dagger}_{\alpha}(\omega')A_{\alpha}(\omega)\right)\right).
    \end{split}
\end{equation}

Since both the dissipator and the Lamb-shift are additive with respect to each bath, we can assume that the correction term to the steady state can be written as a sum of correction terms due separately to each bath. We can then determine the elements $R_{\alpha}(\omega,\omega')$ exactly like in the single bath case. Indeed with the effective detailed balance relation, we can deduce that the eigenoperators and $\rho_{\infty}^{(0)}$ satisfy the following commutation relations 

\begin{equation}
        A_{\alpha}(\omega)\rho^{(0)}_{\infty}=e^{-\tilde\beta(\omega)\omega}\rho^{(0)}_{\infty}A_{\alpha}(\omega).
\end{equation}

Following the same steps as in App.~\ref{appC} we end up with

\begin{equation}
\begin{split}
    &R_{\alpha}(\omega,\omega')=S_{\alpha}(\omega,\omega')+K_{\alpha}(\omega,\omega'),\\
    &K_{\alpha}(\omega,\omega')=-\frac{i\sqrt{\gamma_{\alpha}(\omega)\gamma_{\alpha}(\omega')}}{e^{\tilde\beta(\omega')\omega'-\tilde\beta(\omega)\omega}-1}\left(e^{\tilde\beta(\omega')\omega'-\beta_{\alpha}(\omega+\omega')/2}\right.\\
    &\left.-\frac{(1+e^{\tilde\beta(\omega')\omega'-\tilde\beta(\omega)\omega})}{2}\right).
\end{split}
\end{equation}

\section{Details of the examples}\label{appF}

\subsection{Entropy production analysis for the two qubits}\label{appF1}






The violations appearing in the naive entropy production have been estimated through the calculation of its derivative with respect to the parameter $\mu$ introduced in Eq.\,(\ref{eq:param}). Using properties of the trace of the dissipator, we simplify the computation greatly and get

\begin{align}
\dfrac{d \dot{\sigma}}{d \mu} =&  J (\overline{\omega}) \left[  n(\overline{\omega}) + \dfrac{1}{2} \right] \left[ \bra{01}(\rho +\mu d_{\mu} \rho)\ket{10} \right. \nonumber \\
 &\left.+ \bra{10}(\rho +\mu d_{\mu} \rho)\ket{01} \right]. 
\end{align}

The density matrix was then calculated, along with its trace, its projections and their derivative with respect to $\mu$. The partition function can be simply derived as

\begin{equation}
    Z_{S}(\mu)=\Tr{e^{-\beta \tilde{H}_{S}}} = 1 + e^{-2\beta \bar{\omega}} + 2e^{-\beta \bar{\omega}} \cosh{\beta \Omega(\mu)}.
\end{equation}

Here, $\tilde{H}_{S}$ is the Hamiltonian obtained by absorbing the parameter $\mu$ into the inter-system coupling i. e $g$ is now replaced by $g(\mu) = g - \mu / \beta$ and $\Omega(\mu)=\sqrt{\delta^{2}/4+g(\mu)^{2}}$. The off-diagonal element of the density matrix is given by 

\begin{align}
    \bra{01}\rho\ket{10} & = -\dfrac{g(\mu)e^{-\beta\bar{\omega}}\sinh{ \beta\Omega(\mu)}}{\Omega(\mu)Z_{S}(\mu)}.
\end{align}

We then do an expansion up to second order with respect to $\beta\Omega(\mu)$. We get

\begin{align}
&\dfrac{d}{d\mu}\left[ \bra{01}\rho\ket{10}\right] \approx \left( \dfrac{-2 \beta^{2} g(\mu)^{2} e^{-\beta\bar{\omega}}}{1 + e^{-2\beta \bar{\omega}} + 2e^{-\beta \bar{\omega}}\left( 1 + \dfrac{1}{2}\beta^{2} \Omega(\mu)^{2} \right) } \right.\\
& \left. + \dfrac{\beta^{2} g(\mu)^{2}}{2} + 1\right)\times \dfrac{e^{-\beta\bar{\omega}}}{ 1 + e^{-2\beta \bar{\omega}} + 2e^{-\beta \bar{\omega}} \left( 1 + \dfrac{1}{2}\beta^{2} \Omega(\mu)^{2} \right) }
\end{align}

Assembling everything together and eliminating terms above second order

\begin{equation}
\dfrac{d \dot{\sigma}}{d \mu}\approx \dfrac{J (\overline{\omega})\left( 2 N(\overline{\omega}) + 1 \right)  ( - \beta g(\mu) + \mu  )e^{-\beta\bar{\omega}}}{1 + e^{-2\beta \bar{\omega}} + 2e^{-\beta \bar{\omega}}}
\label{eq:EPR'_approx_th}
\end{equation}

Therefore the minimizer $\mu^{*}$ and the minimum reached by the naive entropy production are respectively given by

\begin{align}
    \mu ^{\ast} &= \dfrac{\beta g}{2} \\
    \dot{\sigma}(\mu^{*})&=-\dfrac{(\beta g)^{2} J(\bar{\omega}) \left( 2N(\bar{\omega}) + 1\right)}{4\cosh^{2}(\beta\overline{\omega}/2)}
\end{align}

\subsection{Dynamics and thermodynamic quantities for the two oscillators}\label{appF2}

To solve the dynamics, we exploit the fact that the total Hamiltonian ($H_{S}+H_{I}+H_{B}$) is gaussian preserving. We can thus fully characterize the state of the system and have access to any observable with its covariance matrix \cite{Weedbrook_2012,Adesso_2014}. In particular, we simply need the knowledge of $\langle a^{\dagger}_{+}a_{+}\rangle$, $\langle a^{\dagger}_{-}a_{-}\rangle$ and $\langle a^{\dagger}_{+}a_{-}\rangle$. From GAME (Eq.~\eqref{GAME_oscillators}), we can deduce the differential equations satisfied by these expectation values and write them in a matrix form 

\begin{equation}
    \frac{dV}{dt}=MV+F,
\end{equation}

with $$V=\left(\langle a^{\dagger}_{+}a_{+}\rangle ,\langle a^{\dagger}_{-}a_{-}\rangle, \Re{\langle a^{\dagger}_{+}a_{-}\rangle},\Im{\langle a^{\dagger}_{+}a_{-}\rangle}\right)^{\intercal},$$ and where M is a 4 by 4 matrix and F a vector. The equation can be formally solved $$V(t)=e^{Mt}(V(0)+M^{-1}F)-M^{-1}F,$$ so that the stationary state is given by $V_\infty=-M^{-1}F$. The heat fluxes can be expressed in terms of the expectation values [see for instance Eq.~\eqref{RightHeat}]. Since the state is Gaussian, the von Neumann entropy of the system can be calculated as the Shannon entropy with respect to the eigenvalues of the covariance matrix of the system \cite{Weedbrook_2012,Adesso_2014}

\begin{equation}
    S=\sum_{i=+,-}(\nu_{i}+1)\ln(\nu_{i}+1)-\nu_{i}\ln(\nu_{i}).
\end{equation}

\end{document}